\documentclass[twocolumn,prl,floatfix,superscriptaddress,nofootinbib,notitlepage]{revtex4-1}

\usepackage{hyperref}
\hypersetup{
	colorlinks = true,
	urlcolor   = blue}
\usepackage{xr}
\usepackage{amsmath}
\usepackage{bm}
\usepackage{gensymb}
\usepackage[normalem]{ulem}
\usepackage{euscript}
\usepackage[pdftex]{graphicx}
\usepackage{xspace}
\usepackage{xfrac}
\usepackage{enumerate}
\usepackage{braket}
\usepackage{float,fancyvrb}
\usepackage[T1]{fontenc}
\usepackage{lmodern}
\graphicspath{{./Figures/}}
\usepackage{wrapfig}
\usepackage{comment}
\usepackage{url}
\usepackage{amssymb}
\usepackage{xcolor} 
\usepackage{lineno} 

\usepackage{mathrsfs}
\usepackage[scr=zapfc, scrscaled=1.15]{mathalfa}

\newcommand{\uu}[1]{\ensuremath{\, \mathrm{#1}}} 
\newcommand{\chem}[1]{\ensuremath{\mathrm{#1}}} 

\let\vec\bm

\definecolor{lilium}{RGB}{146, 111, 202} 
\usepackage[colorinlistoftodos]{todonotes}

\makeatletter
\let\start@align@nopar\start@align
\let\start@gather@nopar\start@gather
\let\start@multline@nopar\start@multline
\long\def\start@align{\par\start@align@nopar}
\long\def\start@gather{\par\start@gather@nopar}
\long\def\start@multline{\par\start@multline@nopar}
\makeatother

\begin{document}
	
	\title{Search for axion-like dark matter using solid-state nuclear magnetic resonance}
	
	\author{Deniz~Aybas}
	\affiliation{Department of Physics, Boston University, Boston, MA 02215, USA}
	\affiliation{Department of Electrical and Computer Engineering, Boston University, Boston, MA 02215, USA}
	\author{Janos~Adam}
	\affiliation{Department of Physics, Boston University, Boston, MA 02215, USA}
	\author{Emmy~Blumenthal}
	\affiliation{Department of Physics, Boston University, Boston, MA 02215, USA}
	\author{Alexander~V.~Gramolin}
	\affiliation{Department of Physics, Boston University, Boston, MA 02215, USA}
	\author{Dorian~Johnson}
	\affiliation{Department of Physics, Boston University, Boston, MA 02215, USA}
	\author{Annalies~Kleyheeg}
	\affiliation{Department of Physics, Boston University, Boston, MA 02215, USA}
	\author{Samer~Afach}
	\affiliation{Helmholtz-Institut, GSI Helmholtzzentrum f{\"u}r Schwerionenforschung, 55128 Mainz, Germany}
	\affiliation{Johannes Gutenberg-Universit{\"a}t Mainz, 55128 Mainz, Germany}
	\author{John~W.~Blanchard}
	\affiliation{Helmholtz-Institut, GSI Helmholtzzentrum f{\"u}r Schwerionenforschung, 55128 Mainz, Germany}
	\author{Gary~P.~Centers}
	\affiliation{Helmholtz-Institut, GSI Helmholtzzentrum f{\"u}r Schwerionenforschung, 55128 Mainz, Germany}
	\affiliation{Johannes Gutenberg-Universit{\"a}t Mainz, 55128 Mainz, Germany}
	\author{Antoine~Garcon}
	\affiliation{Helmholtz-Institut, GSI Helmholtzzentrum f{\"u}r Schwerionenforschung, 55128 Mainz, Germany}
	\affiliation{Johannes Gutenberg-Universit{\"a}t Mainz, 55128 Mainz, Germany}
	\author{Martin~Engler}
	\affiliation{Helmholtz-Institut, GSI Helmholtzzentrum f{\"u}r Schwerionenforschung, 55128 Mainz, Germany}
	\affiliation{Johannes Gutenberg-Universit{\"a}t Mainz, 55128 Mainz, Germany}
	\author{Nataniel~L.~Figueroa}
	\affiliation{Helmholtz-Institut, GSI Helmholtzzentrum f{\"u}r Schwerionenforschung, 55128 Mainz, Germany}
	\affiliation{Johannes Gutenberg-Universit{\"a}t Mainz, 55128 Mainz, Germany}
	\author{Marina~Gil~Sendra}
	\affiliation{Helmholtz-Institut, GSI Helmholtzzentrum f{\"u}r Schwerionenforschung, 55128 Mainz, Germany}
	\affiliation{Johannes Gutenberg-Universit{\"a}t Mainz, 55128 Mainz, Germany}
	\author{Arne~Wickenbrock}
	\affiliation{Helmholtz-Institut, GSI Helmholtzzentrum f{\"u}r Schwerionenforschung, 55128 Mainz, Germany}
	\affiliation{Johannes Gutenberg-Universit{\"a}t Mainz, 55128 Mainz, Germany}
	\author{Matthew~Lawson}
	\affiliation{The Oskar Klein Centre for Cosmoparticle Physics, Department of Physics, Stockholm University, AlbaNova, 10691 Stockholm, Sweden}
	\affiliation{Nordita, KTH Royal Institute of Technology and Stockholm University, Roslagstullsbacken 23, 10691 Stockholm, Sweden}
	\author{Tao~Wang}
	\affiliation{Department of Physics, Princeton University, Princeton, New Jersey, 08544, USA}
	\author{Teng~Wu}
	\affiliation{State Key Laboratory of Advanced Optical Communication Systems and Networks, Department of Electronics, and Center for Quantum Information Technology, Peking University, Beijing 100871, China}
	\author{Haosu~Luo}
	\affiliation{Shanghai Institute of Ceramics, Chinese Academy of Sciences, China}
	\author{Hamdi~Mani}
	\affiliation{School of Earth and Space Exploration, Arizona State University, Tempe, AZ 85287, USA}
	\author{Philip~Mauskopf}
	\affiliation{School of Earth and Space Exploration, Arizona State University, Tempe, AZ 85287, USA}
	\author{Peter~W.~Graham}
	\affiliation{Stanford Institute for Theoretical Physics, Stanford University, Stanford, California 94305, USA}
	\author{Surjeet~Rajendran}
	\affiliation{Department of Physics \& Astronomy, The Johns Hopkins University, Baltimore, Maryland 21218, USA}
	\author{Derek~F.~Jackson~Kimball}
	\affiliation{Department of Physics, California State University - East Bay, Hayward, California 94542-3084, USA}
	\author{Dmitry~Budker}
	\affiliation{Helmholtz-Institut, GSI Helmholtzzentrum f{\"u}r Schwerionenforschung, 55128 Mainz, Germany}
	\affiliation{Johannes Gutenberg-Universit{\"a}t Mainz, 55128 Mainz, Germany}
	\affiliation{Department of Physics, University of California, Berkeley, California 94720-7300, USA}
	\author{Alexander~O.~Sushkov}
	\email{asu@bu.edu}
	\affiliation{Department of Physics, Boston University, Boston, MA 02215, USA}
	\affiliation{Department of Electrical and Computer Engineering, Boston University, Boston, MA 02215, USA}
	\affiliation{Photonics Center, Boston University, Boston, MA 02215, USA}
	
	\date{\today}
	
	\begin{abstract}
We report the results of an experimental search for ultralight axion-like dark matter in the mass range 162 neV to 166 neV. The detection scheme of our Cosmic Axion Spin Precession Experiment (CASPEr) is based on a precision measurement of $^{207}$Pb solid-state nuclear magnetic resonance in a polarized ferroelectric crystal. Axion-like dark matter can exert an oscillating torque on $^{207}$Pb nuclear spins via the electric-dipole moment coupling $g_d$, or via the gradient coupling $g_{\text{aNN}}$. We calibrated the detector and characterized the excitation spectrum and relaxation parameters of the nuclear spin ensemble with pulsed magnetic resonance measurements in a 4.4~T magnetic field. We swept the magnetic field near this value and searched for axion-like dark matter with Compton frequency within a 1~MHz band centered at 39.65~MHz. Our measurements place the upper bounds
	$|g_d|<9.5\times10^{-4}\,\text{GeV}^{-2}$ and
	$|g_{\text{aNN}}|<2.8\times10^{-1}\,\text{GeV}^{-1}$
(95\% confidence level) in this frequency range. The constraint on $g_d$ corresponds to an upper bound of 
	$1.0\times 10^{-21}\,\text{e}\cdot\text{cm}$
on the amplitude of oscillations of the neutron electric dipole moment, and 
	$4.3\times 10^{-6}$
on the amplitude of oscillations of CP-violating $\theta$ parameter of quantum chromodynamics. Our results demonstrate the feasibility of using solid-state nuclear magnetic resonance to search for axion-like dark matter in the nano-electronvolt mass range.
	\end{abstract}
	
	\maketitle
	
	\clearpage
	
	\begin{figure*}[t!]
		\includegraphics[width=0.8\textwidth]{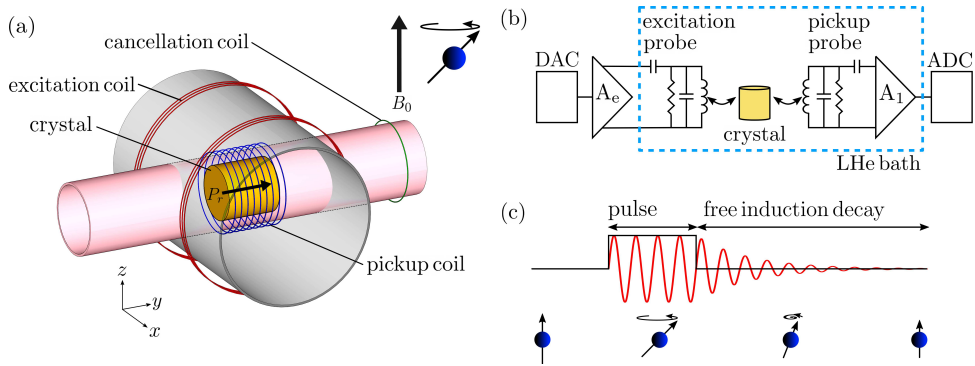}
		\caption{Experimental setup. 
			(a) The sample was a cylindrical ferroelectric PMN-PT crystal with diameter $0.46\uu{cm}$ and thickness $0.50\uu{cm}$. It was electrically polarized along the cylinder axis, indicated with the black arrow. The pickup coil and the cancellation coil were coaxial with the crystal, and the axis of the Helmholtz excitation coil was orthogonal. The vertical leading magnetic field $B_0$ set the direction of the equilibrium spin polarization. Coils were supported by G-10 fiberglass cylinders shown in gray and pink. 
			(b) Electrical schematic, showing the excitation and pickup circuits. Excitation pulses generated with the digital-to-analog converter (DAC) were amplified (A$_e$), and coupled to the excitation coil via a tuned tank circuit that included matching and tuning capacitors, as well as a resistor to set the circuit quality factor. The pickup probe was also designed as a tuned tank circuit, coupling the voltage induced in the pickup coil to a low-noise cryogenic amplifier (A$_1$), whose output was filtered, further amplified, and digitized with an analog-to-digital converter (ADC). 
			(c) Pulsed NMR sequence used for FID measurements. The spin-ensemble equilibrium magnetization, initially parallel to $B_0$, was tilted into the transverse plane by the excitation pulse. The FID signal was recorded after the excitation pulse, as the magnetization precessed and its transverse component decayed.}
		\label{fig:setup}
	\end{figure*}
	
	The existence of dark matter is indicated by astronomical and cosmological evidence, but its interactions, aside from gravity, remain undetected~\cite{Spergel2015,Bertone2018}. A number of theoretical models of physics at high energies, such as string theory, grand unified theories, and models with extra dimensions, incorporate light pseudoscalar bosons (axion-like particles, ALPs), which are potential dark matter candidates~\cite{Preskill1983,Abbott1983,Dine1983,Svrcek2006, Irastorza2018a}. Among these, the axion is particularly compelling, because it also offers a solution to the strong CP problem of quantum chromodynamics (QCD)~\cite{Peccei1977,Weinberg1978,Wilczek1978,DeMille2017,Irastorza2018a}. The axion or axion-like field
	${a(t)=a_0\cos{(\omega_at)}}$ 
	oscillates at the Compton frequency
	${\nu_a=\omega_a/(2\pi)=m_ac^2/h}$, where $c$ is the speed of light in vacuum, $h$ is the Planck constant, and $m_a$ is the unknown ALP mass, which can be in a broad range, roughly between ${10^{-21}\uu{eV}}$ and ${10^{-3}\uu{eV}}$~\cite{Graham2018a,Ernst2018,Schutz2020}. The field amplitude $a_0$ is fixed by the assumption that it dominates the dark matter energy density: 
	${\rho_{\text{DM}} = m_a^2 a_0^2 / 2 \approx 3.6 \times 10^{-42}~\text{GeV}^4}$
	~\cite{PDG2019, Graham2013}. Kinetic energy of the axion-like dark matter field introduces small corrections to its frequency spectrum. The standard halo model predicts the spectral shape with linewidth
	${(v_0^2/c^2)\nu_a\approx 10^{-6}\nu_a}$, where 
	${v_0\approx 220\uu{km/s}}$ is
	the circular rotation speed of the Milky Way galaxy at the Sun's location~\cite{Turner1990,Evans2019}.
	
	Experimental searches for axion-like particles rely on symmetry arguments about the nature of their interactions with Standard Model particles~\cite{Graham2013, Budker2014, Arvanitaki2014, Irastorza2018a}. These interactions are suppressed by a large energy scale, set by the decay constant $f_a$, which could lie near the grand unification, or the Planck scale~\cite{Graham2018}. Most experiments to date have focused on the electromagnetic interaction, which can mix photons with axions and ALPs in the presence of a strong magnetic field~\cite{Sikivie1983,Du2018,Graham2015a, Brubaker2017, Choi2017b,Sikivie2014b, Chaudhuri2015, Kahn2016, Chaudhuri2018, Ouellet2019,Gramolin2020a}. The Cosmic Axion Spin Precession Experiments (CASPEr) search for different interactions: the electric dipole moment (EDM) interaction and the gradient interaction with nuclear spin $\vec{I}$~\cite{Budker2014,Garcon2018,Wang2018,Wu2019a,Garcon2019b,JacksonKimball2020}. The gradient interaction Hamiltonian is 
	${H_{\text{aNN}} = g_{\text{aNN}}\vec{\nabla}a\cdot\vec{I}}$, where $g_{\text{aNN}}$ is the coupling strength.
	The EDM interaction arises from the defining coupling of the axion to the gluon field~\cite{Graham2011}. Its Hamiltonian can be written as 
	${H_{EDM} = g_da\vec{E}^*\cdot\vec{I}/I}$, where $g_d$ is the coupling strength and $E^*$ is an effective electric field~\cite{Budker2014}.
	This interaction is equivalent to that of a parity- and time-reversal-violating oscillating EDM, given by ${d=g_da_0\cos{(\omega_at)}}$.
	This corresponds to an oscillating QCD $\theta$ parameter: ${\theta(t) = (a_0/f_a)\cos{(\omega_at)}}$, with $g_d$ inversely proportional to $f_a$~\cite{Pospelov1999b,Graham2013}.
	The EDM coupling generates axion mass, and for the QCD axion ${m_a \approx \Lambda_{QCD}^2/f_a}$, where ${\Lambda_{QCD}\approx 200\uu{MeV}}$ is the QCD confinement scale~\cite{Baldicchi2007,Graham2013}.
	
	The sensitivity of static EDM experiments to the oscillating EDM is suppressed, although data re-analysis has produced limits at low frequencies~\cite{Abel2017a,Roussy2020}. Astrophysical constraints can be derived by analyzing the cooling dynamics of the supernova SN1987A~\cite{Raffelt2008,Graham2013}. Constraints can also be extracted from analysis of $^4$He production during Big Bang nucleosynthesis~\cite{Blum2014} and from analysis of black hole superradiance~\cite{Arvanitaki2010}. 
	CASPEr-electric is a direct, model-independent search for the EDM and gradient interactions of axion-like dark matter, with the potential to reach the sensitivity to the QCD axion~\cite{Budker2014}. 
	We search for the effects of these interactions on the dynamics of a spin ensemble in a solid with broken inversion symmetry~\cite{Leggett1978,Bialek1986,Mukhamedjanov2005,Budker2006,Sushkov2010,Rushchanskii2010,Eckel2012}.
	The measurements focus on $^{207}$Pb$^{2+}$ ions, with nuclear spin $I=1/2$, in a poled ferroelectric PMN-PT crystal with the chemical formula: 
	\chem{{(PbMg_{1/3}Nb_{2/3}O_3)_{2/3}-(PbTiO_3)_{1/3}}}~\cite{som}.
	The non-centrosymmetric position of the ions in this crystal gives rise to a large effective electric field, analogous to the effect in polar molecules~\cite{Ludlow2013,Skripnikov2016,ACME2018}. The EDM or gradient interaction with axion-like dark matter creates an oscillating torque on the nuclear spins. We quantify the magnitude of this torque by the Rabi frequency $\Omega_a$, which is proportional to the corresponding interaction strength. For a spin ensemble polarized by an external bias magnetic field, this torque tilts the spins, if it is resonant with their Larmor frequency. The experimental observable is the oscillating transverse magnetization:
	\begin{align}
		M_a = uM_0\Omega_aT_2\cos{(\omega_a t)},
		\label{eq:1}
	\end{align}
	where $M_0$ is the equilibrium magnetization of the $^{207}$Pb nuclear spin ensemble, $T_2$ is the nuclear spin coherence time, and $u$ is a dimensionless spectral factor that takes into account the inhomogeneous broadening of the spin ensemble and the detuning between the ALP Compton frequency and the spin Larmor frequency~\cite{som}.
	
	Our apparatus makes use of inductive detection to measure the $^{207}$Pb spin precession, Fig.~\ref{fig:setup}(a). We poled the cylindrical PMN-PT crystal along its axis, aligned with the [1,1,1] crystal direction. This created the axial effective electric field $\vec{E}^*$, proportional to the remanent polarization $P_r$. We mounted the crystal inside a fiberglass tube, so that $\vec{E}^*$ was perpendicular to the vertical bias magnetic field $\vec{B}_0$, created with a superconducting solenoid. A pickup coil, wound around the tube, was coupled to a low-noise cryogenic preamplifier with a tuned matching circuit, Fig.~\ref{fig:setup}(b). We tuned the pickup probe to have its resonance at 39.7~MHz with quality factor $26$, and matched its impedance to the 50~$\Omega$ input impedance of the preamplifier~\cite{som}. A cylindrical copper shield attenuated external sources of RF interference.
	We performed all experiments with the apparatus submerged in a liquid helium bath at $4.2\uu{K}$ temperature~\cite{som}.
	
	We calibrated the pickup probe using $^{207}$Pb pulsed nuclear magnetic resonance (NMR) measurements, Fig.~\ref{fig:setup}(c). The spins were excited by resonant magnetic field pulses, created by delivering current to the $2\times3$-turn Helmholtz excitation coil, coupled to a matching circuit, tuned at $42\uu{MHz}$ with a quality factor 2. The axis of this coil was orthogonal to the pickup coil axis, Fig.~\ref{fig:setup}(a). After each pulse, nuclear spin free induction decay (FID) was measured with the pickup probe, characterized by transfer coefficient ${\alpha=V_1/(\mu_0M_1)}$, where $V_1$ is the recorded voltage referred to the amplifier input, $M_1$ is the transverse sample magnetization, and $\mu_0$ is the permeability of free space. Despite our efforts to minimize the inductive and capacitive couplings between the excitation and the pickup coils, we found that the cryogenic preamplifier saturated during excitation pulses, and its recovery time was too long to observe the fast FID decay~\cite{som}. To address this problem, we placed a single-turn cancellation coil near the pickup coil, Fig.~\ref{fig:setup}(a), and delivered to it a compensating current during the excitation pulses. The amplitude and phase of this compensating current were chosen to cancel the current in the pickup probe during excitation, and prevent preamplifier saturation, without affecting spin excitation. This scheme is a substitute for the transmit/receive switch, often used in NMR detectors. 
	
	\begin{figure*}[t!]
		\includegraphics[width=0.6\textwidth]{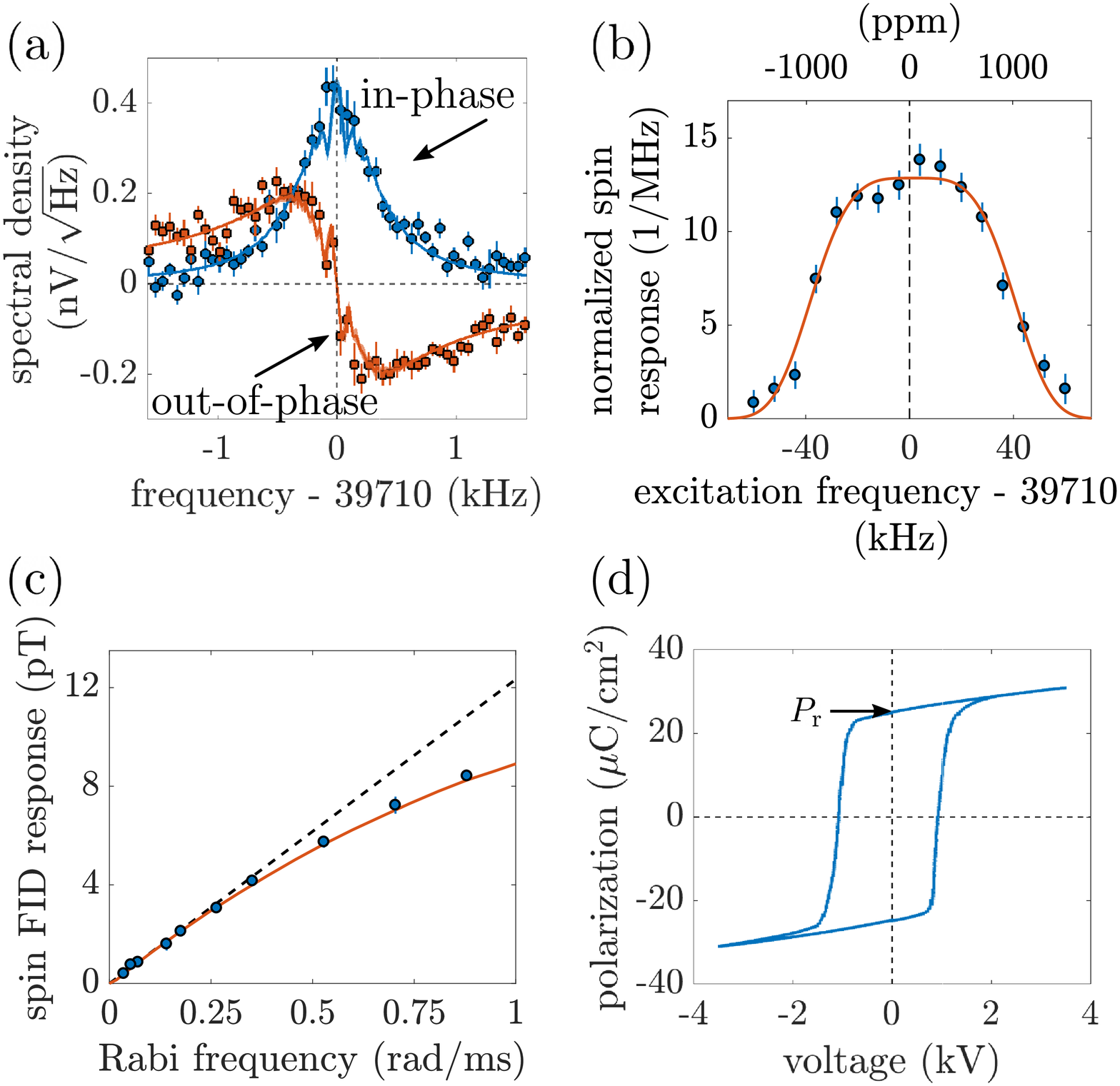}
		\caption{Sensitivity calibration. (a) Measurements of $^{207}$Pb FID following a spin excitation pulse of length ${t_p=20\uu{ms}}$. The excitation carrier frequency was set to ${39.71\uu{MHz}}$, and the Rabi frequency was ${\Omega_e=0.88\uu{rad/ms}}$. The data points show the in-phase (blue circles) and the out-of-phase (orange squares) quadratures of the Fourier transform of the detected voltage, referred to the input of the pickup probe amplifier A$_1$. Data points were binned and averaged, the error bars show one standard deviation for each bin. 
			The lines show the best-fit simulation of the spin response, with the light-colored narrow bands indicating the range of simulation results if parameters are varied by one standard deviation away from their best-fit values.
			We performed the fitting simultaneously to three FID data sets, with excitation pulse lengths $t_p=0.2\uu{ms},\,2\uu{ms},\,20\uu{ms}$, with free parameters including the spin coherence time $T_2$ and pickup circuit transfer coefficient $\alpha$~\cite{som}.
			(b) Measurement of the normalized $^{207}$Pb NMR excitation spectrum near Larmor frequency $39.71\uu{MHz}$. Excitation pulses of length ${1.6\uu{ms}}$ and Rabi frequency ${\Omega_e=0.88\uu{rad/ms}}$ were delivered at the carrier frequencies shown on the x-axis. Data points show the amplitude of the spin FID response, normalized so that the integral of the spectrum is unity.
			The error bars indicate one standard deviation uncertainties of the FID spectrum fits.
			We model the excitation spectrum as a super-Gaussian of order 2 (red line)~\cite{som}.
			(c) Detector calibration for varying drive Rabi frequency. Data points show the amplitude of the spin FID response after an excitation pulse of length 20~ms, delivered at the carrier frequency ${39.71\uu{MHz}}$, with Rabi frequency $\Omega_e$ plotted on the x-axis. 
			The error bars indicate one standard deviation uncertainties, obtained by grouping 100 consecutive FID measurements taken at each $\Omega_e$ into 5 sets, and independently analyzing each set~\cite{som}.
			The orange line shows the spin response simulated using the Bloch equations with parameters extracted from data in panel (a). 
			(d) Measurement of ferroelectric hysteresis in the PMN-PT single crystal. The remanent polarization $P_r$ persists after the applied voltage has been ramped down to zero.
		}
		\label{fig:param}
	\end{figure*}
	
	We performed the NMR calibration measurements at the leading magnetic field ${B_0=4.4\uu{T}}$, for which the value of the equilibrium thermal magnetization $M_0$ of the spin ensemble was ${\mu_0M_0 = 2.9\uu{nT}}$. 
	Before every FID measurement the spin ensemble magnetization was initialized to ${(1.9\pm 0.2)\uu{nT}}$ by saturating the spins, then letting magnetization recover over approximately one population relaxation time~\cite{som}.
	We set the excitation carrier frequency to ${39.71\uu{MHz}}$, and recorded the FID signals after excitation pulses of variable width.
	The Fourier spectrum of one of these FID signals is shown in Fig.~\ref{fig:param}(a).  
	We modeled the FID lineshapes by numerically solving the Bloch equations for a spin ensemble with an inhomogeneously-broadened excitation spectrum~\cite{som}. By fitting the data, we extracted the transverse coherence time of the nuclear spins: ${T_2=(16.7\pm0.9)\uu{ms}}$, and the pickup-circuit transfer coefficient ${\alpha = (2.3\pm0.2)\times10^4\uu{V/T}}$. We note that 
	there is a sharp central feature with linewidth on the order of the Rabi frequency, but the overall FID spectral width is much greater than ${1/T_2}$, since the tilting pulse excites a broad frequency band within the inhomogeneous spin distribution. 
	The exact shape of the FID Fourier spectrum depends on the interplay between the excitation-pulse spectrum, the distribution of tipping angles across the spin ensemble, and the $T_2$ coherence time. 
	
	We measured the inhomogeneous broadening of the $^{207}$Pb nuclear spins in the sample by sweeping the excitation pulse carrier frequency and recording the corresponding FID spectra. The resulting NMR excitation spectrum was centered at ${39.71\uu{MHz}}$ and had a full width ${\Gamma/(2\pi)=(78\pm2)\uu{kHz}}$, Fig~\ref{fig:param}(b). This broadening is consistent with the chemical shift anisotropy (CSA) of $^{207}$Pb observed in solid-state NMR~\cite{Bouchard2008}. We measured the population relaxation time $T_1$ of the $^{207}$Pb nuclear spin ensemble with a saturation-recovery measurement, obtaining ${T_1=(25.8\pm 0.6)\uu{min}}$~\cite{som}.
	
	The spin evolution in our pulsed NMR calibration measurements was more complicated than the CW-like small spin-tip angle response to axion-like dark matter, described by Eq.~(\ref{eq:1}). In order to confirm the validity of our NMR model in the limit of small spin-tip angles, we recorded and analyzed FID data for a range of excitation Rabi frequencies $\Omega_e$. For these measurements we kept the excitation pulse width at ${20\uu{ms}}$ -- approximately the coherence time of axion-like dark matter field with Compton frequency near ${40\uu{MHz}}$. At small excitation amplitudes, the spin response was linear in $\Omega_e$, as described by Eq.~(\ref{eq:1}) for the case of the drive due to interaction with axion-like dark matter, Fig~\ref{fig:param}(c). The slope of the linear response is proportional to the spectral factor ${u=(3.8\pm0.3)\times10^{-4}}$, which is well approximated by the ratio of the homogeneous linewidth ${\pi/T_2}$ and the inhomogeneously-broadened excitation spectrum width $\Gamma$~\cite{som}. The deviation from linearity at larger $\Omega_e$ is due to saturation of the resonant spins in the excitation spectrum, consistent with our Bloch-equation simulations.
	
	Prior to any measurements, the PMN-PT crystal was ferroelectrically poled at room temperature by applying ${3.5\uu{kV}}$ across the crystal faces. We measured the ferroelectric hysteresis loop by sweeping the applied voltage while recording the current flowing through the sample, and integrating it to find the polarization, Fig.~\ref{fig:param}(d). The resulting value of remanent polarization was ${P_r=(22\pm 2)\uu{\mu C/cm^2}}$. We recorded hysteresis data before and after the experiments searching for axion-like dark matter, and verified that the fractional degradation of polarization due to thermal cycling and fatigue was smaller than the quoted uncertainty. The effective electric field $E^*$ is proportional to the ferroelectric polarization~\cite{Mukhamedjanov2005,Ludlow2013,Skripnikov2016}. In order to calculate the value of $E^*$ we considered the Schiff moment $S$ of the $^{207}$Pb nucleus, induced by the oscillating QCD $\theta$ parameter~\cite{Schiff1963,Sandars1967}. The dominant contribution to the Schiff moment arises from the parity- and time-reversal-violating nuclear forces, resulting in the value 
	${S=0.04\theta\uu{e\cdot fm^3}}$ ~\cite{Sushkov1984,Flambaum1986,Khriplovich1997,Flambaum2020a,Yanase2020,som}. This corresponds to the magnitude of effective electric field ${E^*=340\uu{kV/cm}}$. We estimate the theoretical uncertainty in $E^*$ on the level of 50\%~\cite{som}.
	
	\begin{figure*}[t!]
		\includegraphics[width=0.55\textwidth]{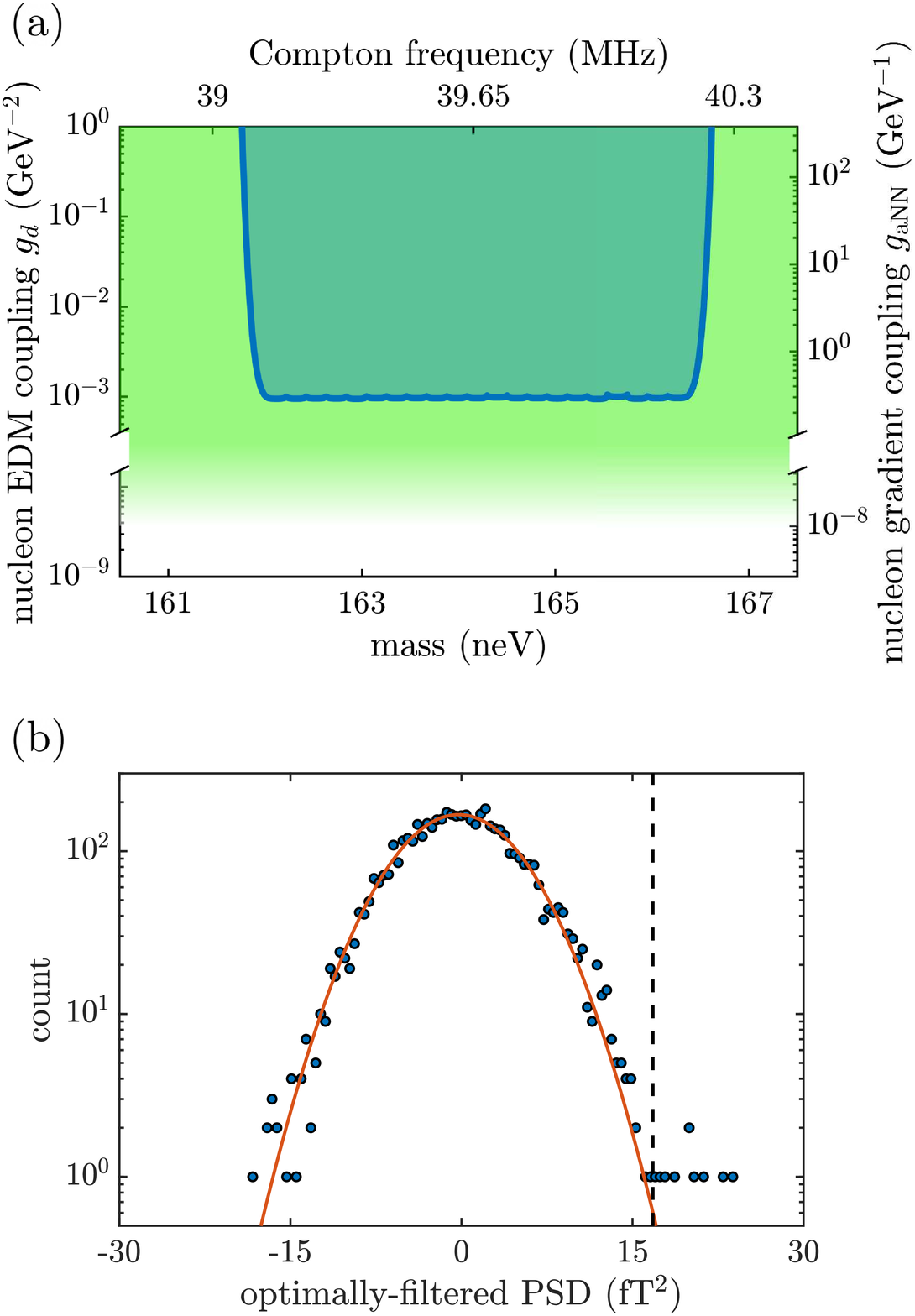}
		\caption{Results of the search for spin interactions with axion-like dark matter. (a) The axion-like dark matter EDM coupling (left y-axis) and nucleon gradient coupling (right y-axis) limits in the mass range $162\uu{neV}-166\uu{neV}$ shown with a blue line.
			The shaded region above the line is excluded at 95\% confidence level. 
			The green region is excluded by analysis of cooling of the supernova SN1987A, the color gradient indicates theoretical uncertainty~\cite{Graham2013}. Existing bounds at other masses, as well as CASPEr sensitivity projections, are shown in Fig.~S9 of the Supplementary Information~\cite{som}.
			(b) The histogram of the optimally-filtered power spectral density of transverse sample magnetization within the frequency window centered at $39.16\uu{MHz}$. The red line shows the Gaussian distribution model, and the vertical black dashed line shows the $3.355\sigma$ candidate threshold at $17\uu{fT^2}$.}
		\label{fig:limit}
	\end{figure*}
	
	In order to search for axion-like dark matter we swept the leading magnetic field $B_0$ in 21 steps, corresponding to the search frequency range ${39.1\uu{MHz}}$ to ${40.2\uu{MHz}}$. The step size was chosen to correspond to ${50\uu{kHz}}$, on the order of the width of the $^{207}$Pb nuclear spin excitation spectrum, Fig.~\ref{fig:param}(b). The broad NMR excitation spectrum reduced the necessary number of magnetic field steps for a given search frequency range. At each value of $B_0$ we recorded 
	${58\uu{s}}$ of scan data sensitive to axion-like dark matter, followed by ${58\uu{s}}$ of re-scan data that were used in our analysis to identify statistical fluctuations. In order to confirm the experimental calibration, we performed pulsed NMR measurements at three values of the leading field, corresponding to the extremes and the midpoint of the search frequency range~\cite{som}.
	
	Data analysis consisted of several processing, correction, and signal-search steps. At each value of the leading field $B_0$ we divided the recorded scan data into $27$ blocks, each of ${2.15\uu{s}}$ duration, chosen to be much longer than the ${\approx 25\uu{ms}}$ coherence time of any potential ALP dark matter signal in our frequency range. We used the pickup-circuit transfer coefficient $\alpha$ to convert the recorded voltage values to magnetization, and performed a discrete Fourier transform on each block, subsequently averaging the power spectral densities (PSDs) of the blocks. 
	Many of the spectra were contaminated with narrowband RF interference that penetrated our electromagnetic shielding. We used Savitzky-Golay digital filtering to identify and reject these narrowband features, while preserving potential axion-like dark matter signals, whose spectral shape is predicted by the standard halo model~\cite{Brubaker2017,Brubaker2017a,som}.
	
	We then processed the data to search for signals due to the EDM and the gradient interactions. 
	The first step was optimal filtering, performed by convolving the PSD with the signal lineshape predicted for the corresponding interaction~\cite{som}. At each value of $B_0$ we retained the optimally-filtered data points in a frequency bin, centered at the corresponding Larmor frequency, with full width ${80\uu{kHz}}$, covering the excitation spectrum bandwidth. We modeled the histogram of these data points as the normal distribution with standard deviation $\sigma$, Fig.~\ref{fig:limit}(b). We set the candidate detection threshold to ${3.355\sigma}$, equivalent to 95\% confidence interval for a ${5\sigma}$ detection, and flagged all points above the threshold as candidates~\cite{Brubaker2017a,Gramolin2020a,som}. 
	
	There were 617 candidates for EDM coupling (636 for gradient coupling). 
	In order to reject residual RF interference, we used the fact that RF pickup is independent of the leading field $B_0$, while an axion-like dark matter signal should only appear when $B_0$ is tuned to a value such that the spin excitation spectrum overlaps with the ALP Compton frequency.
	We compared the candidates from data sets taken at different values of $B_0$, rejecting 569 candidates for EDM coupling (577 for gradient coupling). The remaining 48 candidates for EDM coupling (59 for gradient coupling) were shown to be statistical fluctuations, using a scan/re-scan analysis~\cite{som}. The search sensitivity was limited by the ${\approx0.05\uu{nV/\sqrt{Hz}}}$ input noise level of the amplifier, corresponding to a magnetic field sensitivity of ${\approx2\uu{fT/\sqrt{Hz}}}$.
	
Our search did not yield a discovery of the EDM coupling $g_d$ or the gradient coupling $g_{\text{aNN}}$ of axion-like dark matter. 
In the absence of a detection, in each frequency bin the 95\% confidence interval limit on
magnitudes of these coupling constants corresponds to the $5\sigma$ value in the Gaussian distribution of the optimally-filtered PSD~\cite{Brubaker2017a,Gramolin2020a,som}. 
The limits were corrected to take into account spin saturation~\cite{Castner1959}, normalized by the NMR excitation spectrum for each bin, and concatenated to produce constraints on $g_d$ and $g_{\text{aNN}}$ over the entire frequency search range, Fig.~\ref{fig:limit}(a).
Over the frequency range $39.1\uu{MHz}$ to $40.2\uu{MHz}$ the constraint on $|g_d|$ is 
$|g_d|<9.5\times10^{-4}\,\text{GeV}^{-2}$, corresponding to an upper bound of 
$1.0\times 10^{-21}\,\text{e}\cdot\text{cm}$ on the amplitude of oscillations of the neutron electric dipole moment, and 
$4.3\times 10^{-6}$ on the amplitude of oscillations of the QCD $\theta$ parameter. 
The constraint on $|g_{\text{aNN}}|$ is 
$|g_{\text{aNN}}|<2.8\times10^{-1}\,\text{GeV}^{-1}$. 
The uncertainty on these limits is dominated by the theoretical uncertainty in the effective electric field.
We are not aware of any existing experimental limits on these interactions in this ALP mass range. Analysis of cooling dynamics of supernova SN1987A can be used to estimate bounds $g_d\lesssim10^{-8}\uu{GeV^{-2}}$ and $g_{\text{aNN}}\lesssim10^{-9}\uu{GeV^{-1}}$~\cite{Raffelt2008,Budker2014,Graham2015a}. However these model-dependent bounds are subject to significant caveats and uncertainties, and may be evaded altogether, reinforcing the importance of laboratory searches~\cite{DeRocco2020,Bar2020}. Stringent experimental limits on $g_d$ and $g_{\text{aNN}}$ exist at much lower ALP masses~\cite{Vasilakis2009,Abel2017a,Wu2019a,Adelberger2019,Wu2019b,Garcon2019b,Terrano2019,Roussy2020}.
	
	There are several ways to improve experimental sensitivity to axion-like dark matter. Since the CSA-induced inhomogeneous broadening is proportional to the Larmor frequency, searching in a lower ALP mass range will reduce the linewidth and therefore improve the search sensitivity. A search in the lower mass range will likely also benefit from superconducting detectors, such as SQUIDs and quantum upconverters~\cite{Chaudhuri2019b}. Manipulation of light-induced transient paramagnetic centers may enable control over the nuclear spin population-relaxation time $T_1$, and nuclear spin hyperpolarization using dynamic polarization techniques. A dramatic sensitivity improvement could be achieved by scaling up the sample volume. We estimate that with a sample size of ${\approx 80\uu{cm}}$, it may be possible to reach the sensitivity necessary to detect the QCD axion $g_d$ coupling strength in the mass range between $\approx$~peV and $\approx5$~neV. 
	
	\begin{acknowledgments}
		The authors thank Oyku~Acican for her help with Fig.~\ref{fig:setup}(a), and Alexander~Wilzewski, Hendrik~Bekker, O.~P.~Sushkov, and V.~Flambaum for valuable contributions and discussions. The authors acknowledge support from US Department of Energy grant DESC0019450, the Heising-Simons Foundation grant 2015-039, the Simons Foundation grant 641332, and the Alfred P. Sloan foundation grant FG-2016-6728.
		The work of the Mainz group was supported by the Cluster of Excellence PRISMA+ funded by the German Research Foundation (DFG) within the German Excellence Strategy (Project ID 39083149), by the European Research Council (ERC) under the European Union Horizon 2020 research and innovation program (project Dark-OST, grant agreement No 695405), and by the DFG Reinhart Koselleck project.
		DFJK acknowledges the support of the National Science Foundation under grant PHY-1707875.
	\end{acknowledgments}

\end{document}


\title{Supplemental Material for\\
Search for axion-like dark matter using solid-state nuclear magnetic resonance}

\author{Deniz~Aybas}
\affiliation{Department of Physics, Boston University, Boston, MA 02215, USA}
\affiliation{Department of Electrical and Computer Engineering, Boston University, Boston, MA 02215, USA}
\author{Janos~Adam}
\affiliation{Department of Physics, Boston University, Boston, MA 02215, USA}
\author{Emmy~Blumenthal}
\affiliation{Department of Physics, Boston University, Boston, MA 02215, USA}
\author{Alexander~V.~Gramolin}
\affiliation{Department of Physics, Boston University, Boston, MA 02215, USA}
\author{Dorian~Johnson}
\affiliation{Department of Physics, Boston University, Boston, MA 02215, USA}
\author{Annalies~Kleyheeg}
\affiliation{Department of Physics, Boston University, Boston, MA 02215, USA}
\author{Samer~Afach}
\affiliation{Helmholtz-Institut, GSI Helmholtzzentrum f{\"u}r Schwerionenforschung, 55128 Mainz, Germany}
\affiliation{Johannes Gutenberg-Universit{\"a}t Mainz, 55128 Mainz, Germany}
\author{John~W.~Blanchard}
\affiliation{Helmholtz-Institut, GSI Helmholtzzentrum f{\"u}r Schwerionenforschung, 55128 Mainz, Germany}
\author{Gary~P.~Centers}
\affiliation{Helmholtz-Institut, GSI Helmholtzzentrum f{\"u}r Schwerionenforschung, 55128 Mainz, Germany}
\affiliation{Johannes Gutenberg-Universit{\"a}t Mainz, 55128 Mainz, Germany}
\author{Antoine~Garcon}
\affiliation{Helmholtz-Institut, GSI Helmholtzzentrum f{\"u}r Schwerionenforschung, 55128 Mainz, Germany}
\affiliation{Johannes Gutenberg-Universit{\"a}t Mainz, 55128 Mainz, Germany}
\author{Martin~Engler}
\affiliation{Helmholtz-Institut, GSI Helmholtzzentrum f{\"u}r Schwerionenforschung, 55128 Mainz, Germany}
\affiliation{Johannes Gutenberg-Universit{\"a}t Mainz, 55128 Mainz, Germany}
\author{Nataniel~L.~Figueroa}
\affiliation{Helmholtz-Institut, GSI Helmholtzzentrum f{\"u}r Schwerionenforschung, 55128 Mainz, Germany}
\affiliation{Johannes Gutenberg-Universit{\"a}t Mainz, 55128 Mainz, Germany}
\author{Marina~Gil~Sendra}
\affiliation{Helmholtz-Institut, GSI Helmholtzzentrum f{\"u}r Schwerionenforschung, 55128 Mainz, Germany}
\affiliation{Johannes Gutenberg-Universit{\"a}t Mainz, 55128 Mainz, Germany}
\author{Arne~Wickenbrock}
\affiliation{Helmholtz-Institut, GSI Helmholtzzentrum f{\"u}r Schwerionenforschung, 55128 Mainz, Germany}
\affiliation{Johannes Gutenberg-Universit{\"a}t Mainz, 55128 Mainz, Germany}
\author{Matthew~Lawson}
\affiliation{The Oskar Klein Centre for Cosmoparticle Physics, Department of Physics, Stockholm University, AlbaNova, 10691 Stockholm, Sweden}
\affiliation{Nordita, KTH Royal Institute of Technology and Stockholm University, Roslagstullsbacken 23, 10691 Stockholm, Sweden}
\author{Tao~Wang}
\affiliation{Department of Physics, Princeton University, Princeton, New Jersey, 08544, USA}
\author{Teng~Wu}
\affiliation{State Key Laboratory of Advanced Optical Communication Systems and Networks, Department of Electronics, and Center for Quantum Information Technology, Peking University, Beijing 100871, China}
\author{Haosu~Luo}
\affiliation{Shanghai Institute of Ceramics, Chinese Academy of Sciences, China}
\author{Hamdi~Mani}
\affiliation{School of Earth and Space Exploration, Arizona State University, Tempe, AZ 85287, USA}
\author{Philip~Mauskopf}
\affiliation{School of Earth and Space Exploration, Arizona State University, Tempe, AZ 85287, USA}
\author{Peter~W.~Graham}
\affiliation{Stanford Institute for Theoretical Physics, Stanford University, Stanford, California 94305, USA}
\author{Surjeet~Rajendran}
\affiliation{Department of Physics \& Astronomy, The Johns Hopkins University, Baltimore, Maryland 21218, USA}
\author{Derek~F.~Jackson~Kimball}
\affiliation{Department of Physics, California State University - East Bay, Hayward, California 94542-3084, USA}
\author{Dmitry~Budker}
\affiliation{Helmholtz-Institut, GSI Helmholtzzentrum f{\"u}r Schwerionenforschung, 55128 Mainz, Germany}
\affiliation{Johannes Gutenberg-Universit{\"a}t Mainz, 55128 Mainz, Germany}
\affiliation{Department of Physics, University of California, Berkeley, California 94720-7300, USA}
\author{Alexander~O.~Sushkov}
\email{asu@bu.edu}
\affiliation{Department of Physics, Boston University, Boston, MA 02215, USA}
\affiliation{Department of Electrical and Computer Engineering, Boston University, Boston, MA 02215, USA}
\affiliation{Photonics Center, Boston University, Boston, MA 02215, USA}

\date{\today}

\maketitle

\hypersetup{linkcolor=blue}

\onecolumngrid 

\section{Experimental setup}

\subsection{Description of the apparatus}
\begin{figure}[h]
	\includegraphics[width=0.5\textwidth]{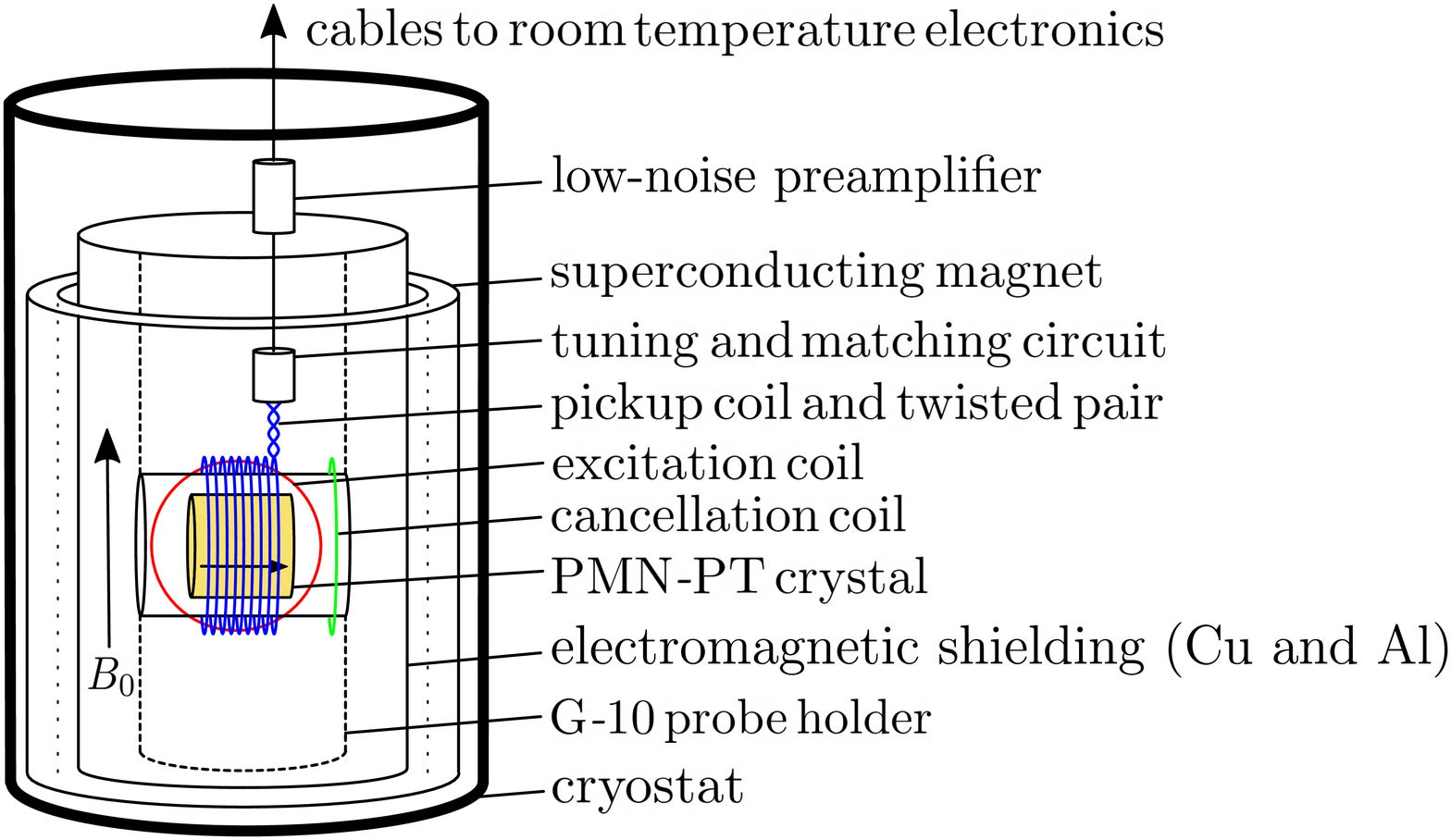}
	\caption{Schematic of the setup operating at 4.2~K. Cylindrical PMN-PT crystal is placed close to the center of the superconducting magnet, and it is coupled to the mutually orthogonal excitation and pickup coils. The sample and the pickup probe are mounted inside a cylindrical electromagnetic shielding enclosure within the superconducting magnet bore. The low-noise preamplifier is inside the liquid helium bath, above the superconducting magnet.}
	\label{fig:setup1}
\end{figure}

\noindent
Our cryogenic nuclear magnetic resonance (NMR) setup is inside a liquid helium (LHe) bath cryostat with a solenoidal superconducting magnet (Cryomagnetics, Inc. Model 90-300-010), Fig.~\ref{fig:setup1}. The apparatus is built around a crystal that is inductively coupled to a pickup probe along one axis, and an excitation probe along an orthogonal axis, both in the plane transverse to the leading magnetic field created by the magnet (Fig. 1(a) in the main text). The experimental setup is used both when measuring pulsed NMR and when performing the axion search.

\begin{figure}[h]
	\includegraphics[width=1\textwidth]{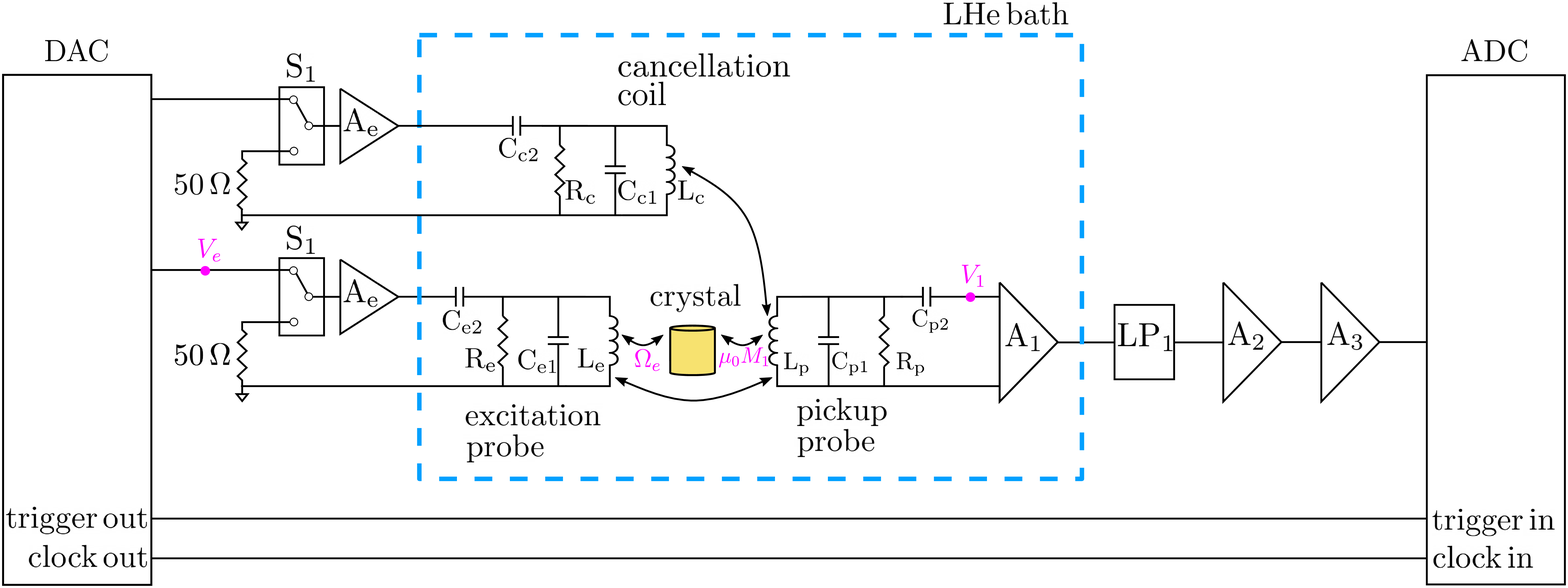}
	\caption{Full electrical schematic of the experimental setup incorporating the Spectrum Instrumentation m4i.6622-x8 digital-to-analog converter (DAC), RF Lambda coaxial reflective  SP2T RFSP2TRDC06G switches ($S_1$), Stanford Research Systems SIM954 inverting amplifiers ($A_e$), coil inductances ($L_c$, $L_e$ and $L_p$), surface mount probe tuning capacitors ($C_{c1}$, $C_{e1}$ and $C_{p1}$), surface mount impedance matching capacitors ($C_{c2}$, $C_{e2}$ and $C_{p2}$), surface mount resistors  ($R_{c}$, $R_{e}$ and $R_{p}$) that determine the quality factor of the circuit resonances, ULF-LNA-$\#$159 cryogenic low-noise preamplifier designed and constructed by the Arizona State University group ($A_1$), Mini-Circuits ZX75LP-50-S+ 50 MHz low-pass filter ($LP_1$), Mini-Circuits ZX60-P103LN+ low-noise amplifier ($A_2$), Femto HVA-200M-40-B amplifier ($A_3$), and Spectrum Instrumentation m4i.4421-x8 analog-to-digital converter (ADC).}
	\label{fig:setup2}
\end{figure}

During pulsed NMR calibration measurements, a digital-to-analog converter (DAC) generates a radio frequency (RF) voltage waveform $V_e$, which is coupled into the excitation probe (Fig.~\ref{fig:setup2}). The resulting RF magnetic field exerts a torque on the spins, whose magnitude is quantified by the excitation Rabi frequency $\Omega_e$. The excitation-probe transfer function is defined as
\begin{align}
	\kappa = \frac{\Omega_e}{V_e}.
	\label{eq:010}
\end{align}

The excitation pulse tilts $^{207}\mathrm{Pb}$ nuclear spins into the plane transverse to the leading field $B_0$, creating a crystal magnetization $M_1$ that rotates at the Larmor frequency. After the excitation pulse ends, this magnetization decays (free induction decay, FID). The magnetization induces an oscillating current in the pickup coil, and voltage $V_1$ at the input of the low-noise preamplifier $A_1$ (Fig.~\ref{fig:setup2}). The pickup probe transfer function is defined as
\begin{align}
	\alpha = \frac{V_1}{\mu_0 M_1},
	\label{eq:020}
\end{align}
where $\mu_0$ is the permeability of free space.
The preamplifier $A_1$ has gain of $40$. Its output is connected to a low-pass filter $LP_1$ and another amplifier stage $A_2$ (gain $= 15$) mounted inside the cryostat near the top flange. After a third amplifier stage $A_3$ (gain $= 10$) outside the cryostat, the signal is sent to an analog-to-digital converter (ADC). 

The excitation signal is routed through a switch $S_1$ (Fig.~\ref{fig:setup2}) that is controlled with a transistor-transistor logic (TTL) pulse with the same duration as the excitation RF pulse. This prevents the DAC output noise from coupling into the pickup probe after the end of the excitation pulse, during FID detection. When the TTL state is high at $5\uu{V}$, the DAC is connected to the excitation probe through amplifier $A_e$, and when the TTL state is low at $0\uu{V}$ the input of $A_e$ is connected to ground via a $50\,\Omega$ termination.

\subsection{The crosstalk minimization scheme}
\noindent
During experimental assembly we carefully adjust the orthogonal axes of the excitation and the pickup coils to minimize mutual inductance between them, to $\approx 1.5\%$ of its maximum value for parallel axes.
Despite these efforts, the excitation pulse induces a crosstalk current in the pickup coil, with amplitude and phase depending on the residual inductive and capacitive couplings between the coils. This crosstalk signal saturates the preamplifier, resulting in a recovery time of~$\approx200\uu{\mu s}$, which complicates the detection of the FID signal. In order to prevent saturation, during the excitation pulse we apply a waveform to the cancellation coil that is optimized to compensate the crosstalk current in the pickup probe with minimal effect on spin dynamics. The phase and amplitude of this waveform are optimized by monitoring the current measured at the pickup probe and minimizing its magnitude. This is done at zero leading magnetic field to avoid spin excitation during optimization. We emphasize that only a small ($<1.5\%$) fraction of the excitation pulse RF field couples into the pickup probe and has to be compensated, therefore our compensation scheme has a correspondingly small effect on spin dynamics.

In many room-temperature NMR measurements, preamplifier saturation is prevented by using a transmit/receive (T/R) switch between the pickup probe and the preamplifier. Because our preamplifier is at $4.2\uu{K}$ temperature, we chose to use the compensation scheme discussed above, rather than designing and constructing a cryogenic T/R switch.

\subsection{Tuning and matching of pickup probe, excitation probe, and cancellation coil}
\noindent
Our magnetic resonance probes are designed as series capacitance-tuned tank circuits, Fig.~\ref{fig:setup2}. In these circuits, coil inductance $L$ is in parallel with a tuning capacitor $C_1$ and a resistor $R$, and this tank circuit is in turn in series with a matching capacitor $C_2$.
The total probe impedance is
\begin{align}
	Z &= \frac{1}{\frac{1}{i\omega L} + \frac{1}{R} + i\omega C_1} +  \frac{1}{i\omega C_2} \nonumber\\
	&= \left(\frac{(\omega L)^2 R}{R^2(1-\omega^2 L C_1)^2 + (\omega L)^2}\right) + i \left(\frac{\omega L R^2 (1-\omega^2 L C_1)}{R^2 (1-\omega^2 L C_1)^2 + (\omega L)^2}  - \frac{1}{\omega C_2}\right).
\end{align}
In order to match the probe impedance to $Z=R_0=50\uu{\Omega}$ at the resonance angular frequency $\omega_r$, we have to choose the following values for the circuit elements:
\begin{align}
	R &= Q \omega_r L \nonumber, \\
	C_1 &= \frac{1}{\omega_r^2 L} \left( 1 - \frac{1}{Q}\sqrt{\frac{R-R_0}{R_0}} \right), \\
	C_2 &= \frac{1}{\omega_r} \sqrt{\frac{1}{R_0(R-R_0)}}, \nonumber 
\end{align}
where $Q$ is the resonance quality factor. We used fixed-value surface mount capacitors and resistors, so the probes are not tunable after the setup is assembled.

The pickup coil with $N_p = 9$ turns of 26 AWG (American Wire Gauge) copper wire is a solenoid with return path cancellation. It has a radius $r_p = 3.2\uu{mm}$, an inductance of $L_p = 0.5 \uu{\mu H}$, and is tuned to the resonant frequency $\omega_p/(2\pi) = 39.71 \uu{MHz}$ with quality factor $Q_p = 26$ at $4.2\uu{K}$. The width of the pickup probe resonance limits the frequency range over which we can search for axion-like dark matter without re-tuning the probe. This is why we limited $Q_p$ to 26. The excitation coil has a Helmholtz geometry with $N_e = 2\times3$ turns of $26$ AWG copper wire with radius $r_e = 7.1\uu{mm}$ and inductance of $L_e = 0.3\uu{\mu H}$, which is tuned to the resonant frequency $\omega_e/(2\pi) = 42.01 \uu{MHz}$ with quality factor $Q_e = 1.5$ at $4.2\uu{K}$. The cancellation coil is a single turn loop around $r_c=4.8\uu{mm}$ radius with $26$ AWG copper and inductance of $L_c = 0.01 \uu{\mu H}$, which is tuned to resonant frequency $\omega_c/(2\pi) = 40.31 \uu{MHz}$ with quality factor $Q_c = 2$ at $4.2\uu{K}$. All the probes are matched to $50\uu{\Omega}$.

\subsection{Estimates of the pickup probe transfer function $\alpha$ and the excitation probe transfer function $\kappa$}\label{sec:D}
\noindent
Based on the electrical schematic described above, let us estimate the values of the transfer functions $\alpha$ and $\kappa$, defined in Eqs.~(\ref{eq:010},\ref{eq:020}). Using Faraday's law we can estimate the voltage induced in the pickup coil by an oscillating transverse magnetization $M_1$:
\begin{align}
	V_p \approx \frac{1}{3} (\gamma B_0) N_p (\mu_0M_1) (\pi r_{s}^2),
\end{align}
where $B_0=4.4\uu{T}$ is the leading static magnetic field, $r_s = 2.3\uu{mm}$ is the radius of the sample, and $\gamma$ is the $^{207}$Pb gyromagnetic ratio. We set the demagnetizing factor to $1/3$, as for a sphere, as an approximation for a cylindrical sample with height $\approx$ diameter. For the pickup probe on resonance with the spin Larmor frequency, the resulting voltage at the input of preamplifier $A_1$ is calculated from circuit analysis~\cite{Miller2000}:
\begin{align}
	V_1 \approx \frac{V_p}{2} \sqrt{\frac{Q_pR_0}{\omega_p L_p}},
\end{align}
while $Q_p \omega_p L_p \gg R_0$. We can therefore estimate the pickup probe transfer function:
\begin{align}
	\alpha = \frac{V_1}{\mu_0M_1} \approx \frac{1}{3} (\gamma B_0)N_p (\pi r_{s}^2) \left(\frac{1}{2} \sqrt{\frac{Q_pR_0}{\omega_p L_p}}\right)
	\approx 2\times10^{4}\uu{V/T}.
	\label{eqn:alpha}
\end{align}

For an excitation voltage $V_e$, referred to the output of the DAC, the current through the excitation coil is calculated from circuit analysis:
\begin{align}
	I_e \approx A_e V_e {\sqrt{\frac{Q_e}{R_0 \omega_e L_e}}},
\end{align}
where $|A_e| = 4$ is the gain of the SRS SIM954 amplifier. Note that the SIM954 has an output impedance $3.3\uu{\Omega}\ll 50\uu{\Omega}$.
The magnetic field produced by this current at the center of the Helmholtz excitation coil is 
\begin{align}
	B_e = \mu_0 I_e \frac{r_e^2}{\left(\sqrt{(r_e/2)^2+r_e^2}\right)^3} \frac{N_e}{2}.
\end{align}
Assuming the excitation is resonant with the spin Larmor frequency, the Rabi frequency is $\Omega_e = \gamma (B_e/2)$. The factor of $1/2$ arises because only  one circular component of the linearly polarized excitation magnetic field $B_e$ is resonant (rotating wave approximation). Therefore the excitation probe transfer function can be estimated as
\begin{align}
	\kappa = \frac{\Omega_e}{V_e} \approx \frac{\gamma \mu_0}{2}  \left(A_e{\sqrt{\frac{Q_e}{R_0 \omega_e L_e}}}\right) \frac{r_e^2}{\left(\sqrt{(r_e/2)^2+r_e^2}\right)^3} \frac{N_e}{2}
	\approx 0.4\uu{rad/(ms\cdot V)} .
	\label{eqn:kappa}
\end{align}

Section \ref{simbloch} describes how we used pulsed NMR to measure the values of $\alpha$ and $\kappa$. The proximity of the measured values to the estimates above validates the approximations used when analyzing the apparatus design shown in Fig.~\ref{fig:setup2}. 

\subsection{Shielding to reduce RF interference}
\noindent
The probes are mounted on a G-10 fiberglass cylinder, with a~0.2-mm thick copper sheet wrapped around the outside. The cylinder is positioned inside the magnet bore, Fig.~\ref{fig:setup1}. The copper sheet forms a closed shielding enclosure, together with aluminum end caps on top and bottom. The RG316 coaxial cable between the pickup probe and the low-noise amplifier $A_1$ is shielded with a~0.5-mm thick copper mesh sleeve. Another copper mesh sleeve shields the bundled RG316 coaxial cables that run up to the top flange of the cryostat. Shields are connected to the ground pin of the $A_1$ amplifier used as a common ground. We estimate the RF interference noise reduction factor due to the shields to be on the order of~$10$ within the $1\uu{MHz}$ range centered at~$39.71\uu{MHz}$.

\section{Characterization of the setup with nuclear magnetic resonance}
\subsection{Properties of the $^{207}$Pb spin ensemble}
\noindent
The $^{207}$Pb isotope has nuclear spin $I=1/2$ and gyromagnetic ratio
\begin{align}
	\frac{\gamma}{2\pi} = \frac{\mu}{h I} = 9.03\uu{MHz/T},
\end{align}
where $\mu = 0.5926\mu_{\mathrm{N}}$ is the magnetic moment of $^{207}\mathrm{Pb}$ nucleus~\cite{NIST2013}, and the nuclear magneton is $\mu_N/h= 7.6226\uu{MHz/T}$~\cite{CODATA2014}.

The chemical formula of PMN-PT is \chem{(PbMg_{1/3}Nb_{2/3}O_3)_{2/3}-(PbTiO_3)_{1/3}}. The number density of $^{207}$Pb nuclear spins in a PMN-PT crystal is given by:
\begin{align}
	n = \frac{\rho}{M} \cdot N_A \cdot 0.221 = 3.4\times10^{27}\uu{m^{-3}},
\end{align}
where $\rho = 8.2\uu{g/cm^{3}}$~\cite{Kochary2007} is the mass density, $M=317.9\uu{g/mole}$~\cite{NIST2013} is the molar mass, and $N_A$ is the Avogadro constant. The natural abundance of $^{207}\mathrm{Pb}$ is $22.1\%$~\cite{NIST2013}.

We perform our experiments in the leading magnetic field $B_0=4.4\uu{T}$ and at temperature $T=4.2\uu{K}$. The equilibrium magnetization of the $^{207}\mathrm{Pb}$ nuclear spin ensemble is given by~\cite{Abragam1961}
\begin{align}
	\mu_0M_0 = \mu_0 \frac{n\gamma^2\hbar^2I(I+1)B_0}{3k_BT}=2.9\uu{nT},
	\label{eq:M0}
\end{align}
where $k_B$ is the Boltzmann constant, $\mu_0$ is the permeability of free space, and $\hbar$ is the reduced Planck constant.

We model the NMR excitation spectrum as a super-Gaussian distribution of order 2, given by
\begin{align}
	f(\nu) = \frac{6.33}{\Gamma}\exp{\left(-\left[\frac{2(\nu-\nu_0)}{\Gamma/(2\pi)}\right]^4\ln{2}\right)},
	\label{eqn:gamma}
\end{align}
where $\Gamma/(2\pi)$ is the full-width at half-maximum, $\nu$ is the excitation frequency, and $\nu_0$ is the center of the distribution.
The scaling pre-factor is chosen to ensure that the area under the distribution is normalized to 1.

\subsection{Saturation-recovery measurements of the relaxation time $T_1$}\label{sec:satur}
\begin{figure}[h]
	\includegraphics[width=0.35\textwidth]{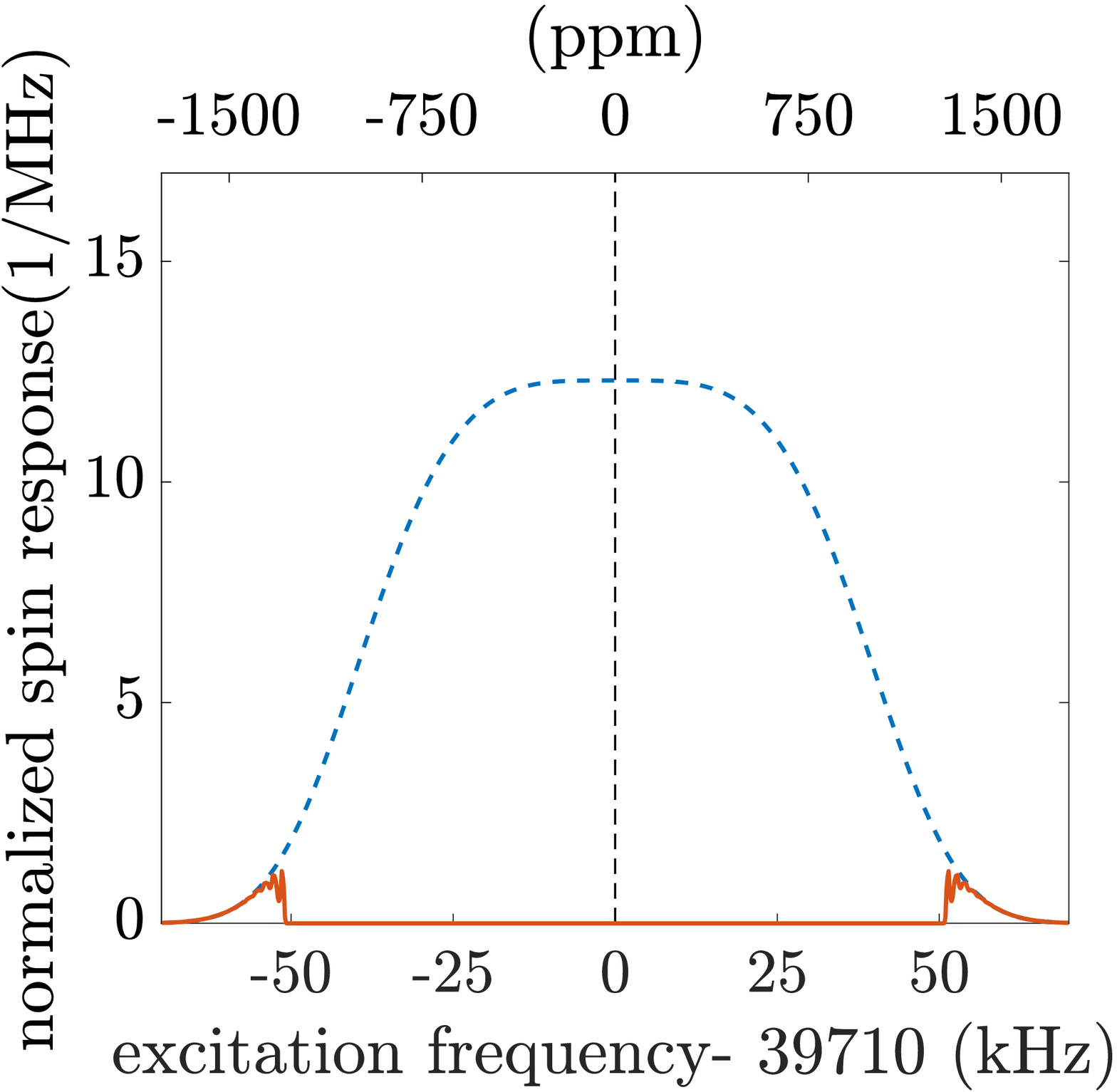}
	\caption{Saturation of the spin-ensemble excitation spectrum by the sequence described in section~\ref{sec:satur}. Bloch equation simulations are performed as described in section~\ref{simbloch}. The initial spin excitation spectrum is shown by the blue dashed line. The spectrum immediately after the saturation step is shown by the orange solid line.}
	\label{fig:saturation}
\end{figure}

\noindent
We use the standard NMR saturation recovery scheme to measure the $T_1$ relaxation time of the $^{207}$Pb nuclear spin ensemble in PMN-PT at $4.2\uu{K}$. Each measurement begins with a saturation step, comprising $100$ consecutive repetitions of a sequence with $101$ pulses whose carrier frequencies vary across the width of the excitation spectrum from $39.66\uu{MHz}$ to $39.76\uu{MHz}$, and whose Rabi frequencies are fixed at $0.88\uu{rad/ms}$. Each pulse duration is $0.8\uu{ms}$, and the pulse spacing is $1.4\uu{ms}$. Bloch-equation simulations confirm that this step saturates the spin ensemble, Fig.~\ref{fig:saturation}.

The saturation step is followed by a variable recovery wait time $t$, after which a pulsed NMR measurement is performed, with spin FID recorded after excitation pulses of $20\uu{ms}$ duration and $180\uu{ms}$ repetition time. The dependence of the FID amplitude on recovery time $t$ is modeled as an exponential $1-e^{-t/T_1}$. The best-fit value for the population relaxation time is $T_1=(25.8\pm 0.6)\uu{min}$.

\subsection{Spin-dynamics simulations with Bloch equations}\label{simbloch}
\noindent
We use the Bloch equations to quantitatively describe the magnetic resonance dynamics of the $^{207}$Pb nuclear spin ensemble~\cite{Bloch1946,Abragam1961}. We choose the direction of the z-axis to be along the static magnetic field $B_0$. The linearly-polarized excitation magnetic field $B_e = (2\Omega_e/\gamma) \cos{(\omega_1t)}$ is applied in the $x$-direction. In the reference frame that rotates at the angular frequency $\omega_1$ around the leading magnetic field, the Bloch equations read
\begin{align}
	\frac{\d\tilde{M}_x}{\d t} &= - \frac{\tilde{M}_x}{T_2} + \Delta\omega\tilde{M}_y ,\nonumber\\
	\frac{\d\tilde{M}_y}{\d t} &= - \Delta\omega\tilde{M}_x - \frac{\tilde{M}_y}{T_2} - \Omega_e M_z ,\\
	\frac{\d M_z}{\d t} &= \Omega_e \tilde{M}_y -\frac{M_z - M'_0}{T_1} ,\nonumber
\end{align}
where 
$\Delta\omega=\omega_1-\omega_0$
is the detuning of spin Larmor frequency $\omega_0$ from the rotating frame frequency, $M'_0$ is the initial ensemble magnetization, $T_2$ is the transverse spin coherence time, and $\tilde{M}_{x,y}$ are the transverse spin magnetization components in the rotating frame. The transformation between magnetization in the laboratory and the rotating frames is
\begin{align}
	M_x = \tilde{M}_{x} \cos(\omega_1t) - \tilde{M}_{y} \sin(\omega_1t) ,\nonumber\\
	M_y = \tilde{M}_{x} \sin(\omega_1t) + \tilde{M}_{y} \cos(\omega_1t),
\end{align}
where in the lab frame $\vec{M} = M_x\hat{x} + M_y \hat{y} + M_z \hat{z}$. 

We numerically solve the Bloch equations using the Runge-Kutta method. The inhomogeneously-broadened spin ensemble is represented by $3251$ spins, with their Larmor frequencies uniformly distributed in an excitation bandwidth of $65\uu{kHz}$ with $0.02\uu{kHz}$ spacing. We simulate the dynamics of each spin independently, and add their contributions to obtain the total magnetization. 

\begin{figure}[h]
	\includegraphics[width=0.6\textwidth]{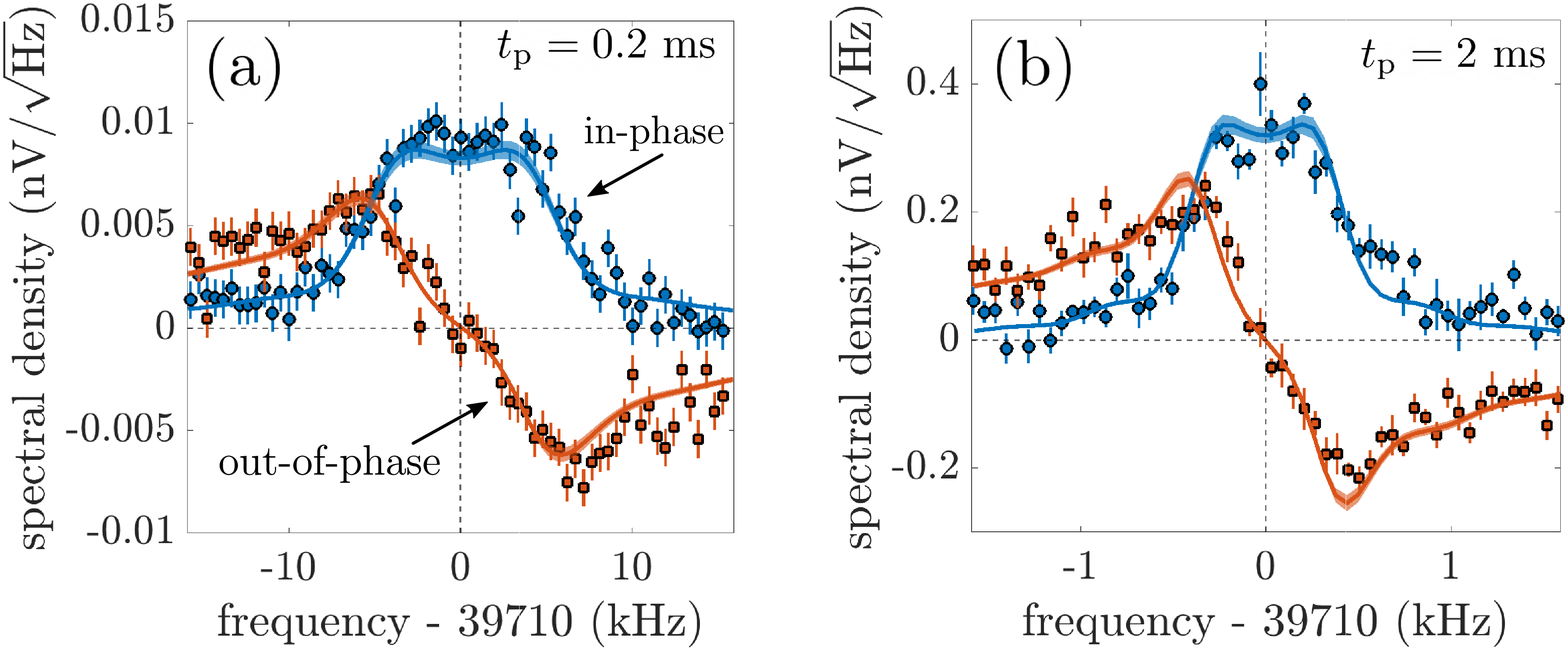}
	\caption{Measurements of $^{207}$Pb FID spectra following a spin excitation pulse of length $t_p$, as indicated in the panels. We performed fitting simultaneously to in-phase (blue) and out-of-phase (orange) components of Fourier transforms of averaged FID from three data sets: with excitation pulse duration $t_p=20\uu{ms}$ shown in Fig. 2(a) in the main text, (a) with excitation pulse duration $t_p=0.2\uu{ms}$, and (b) with excitation pulse duration $t_p=2\uu{ms}$. Data points were binned, and the error bars show one standard deviation in each bin. 
		The lines show the best-fit simulation of the spin response, with the light-colored narrow bands indicating the range of simulation results if parameters $T_2$, $\kappa$, and $\alpha$ are varied by one standard deviation away from their best-fit values. We note that there is a sharp central feature with linewidth on the order of the Rabi frequency, visible in the FID spectrum shown in Fig. 2(a) of main text. Simulations show that the shapes of the FID Fourier spectra depend on the interplay between the excitation-pulse spectrum, the distribution of tipping angles across the spin ensemble, and the $T_2$ coherence time.
	}
	\label{fig:nmr}
\end{figure}

The simulation parameters are the spin coherence time $T_2$, and the transfer functions $\alpha$ and $\kappa$, defined in Sec.~\ref{sec:D}. We perform fits to experimental FID spectra, shown in Fig. 2(a) of the main text and Fig.~\ref{fig:nmr}, by varying the values of these parameters to achieve the minimum value of the goodness-of-fit parameter $\chi^2 = \chi^2_1+\chi^2_2+\chi^2_3$, where the subscript enumerates the measurements with different pulse duration $t_p = 0.2\uu{ms},\,2\uu{ms},\,20\uu{ms}$. For each measurement $i=1,2,3$

\begin{align}
	\chi_i^2 = \sum_{\nu} \left[\operatorname{Re}\left(F_{\text{exp}}[\nu]-F_{\text{sim}}[\nu]\right)^2 + \operatorname{Im}\left(F_{\text{exp}}[\nu]-F_{\text{sim}}[\nu]\right)^2\right],
\end{align}
where $F_{exp}$ is the Fourier transform of the experimentally detected voltage and $F_{sim}$ is the Fourier transform of simulation results, converted into voltage using the transfer coefficient $\alpha$, and the index $\nu$ labels discrete frequency points within the window shown in Fig. 2(a) of the main text and Fig.~\ref{fig:nmr}. The real part of the Fourier transform corresponds to the in-phase quadrature, and the imaginary part corresponds to the out-of-phase quadrature of the FID, relative to the carrier phase of the excitation pulse. 

The excitation pulses induce probe ringing with time constant $\approx 500\uu{ns}$, therefore we use the FID response data starting at $5\uu{\mu s}$ after the end of an excitation pulse. To improve the signal-to-noise ratio, we average the recorded FID response data for several consecutive excitation pulses: 10 data sets are averaged for $t_p=0.2\uu{ms}$ pulse duration, 4 data sets are averaged for $t_p=2\uu{ms}$ pulse duration, and 4 data sets are averaged for $t_p=20\uu{ms}$ pulse duration. After performing the discrete Fourier transform, data points are binned along the frequency axis, with 4 points per bin for $t_p=0.2\uu{ms}$ pulse duration, 2 points per bin for $t_p=2\uu{ms}$ pulse duration, and 2 points per bin for $t_p=20\uu{ms}$ pulse duration. The error bars shown in Fig. 2(a) of the main text and Fig.~\ref{fig:nmr} are the standard deviation of the points within each bin.

The spin ensemble was saturated before every FID measurement, and the FID measurements started after a wait time $\approx T_1$ after saturation. 
Therefore the initial magnetization at the start of every FID measurement was $\mu_0M'_0 = (0.67\pm0.05)\mu_0M_0 = (1.9\pm0.2)\uu{nT}$, where $M_0$ is the thermal equilibrium ensemble magnetization given by Eq.~(\ref{eq:M0}).

Using the measurements shown in Fig. 2(a) in the main text and Fig.~\ref{fig:nmr}, we extract the best-fit parameter values: 
\begin{align}
	T_2 &= (16.7\pm0.9)\uu{ms},\nonumber\\
	\kappa &= (0.352\pm0.007)\uu{rad/(ms \cdot V)},\\
	\alpha &= (2.3\pm0.2)\times10^4\uu{V/T}.\nonumber
\end{align}
The uncertainties are evaluated by bootstrapping: the frequency-domain data are down-sampled into 16 groups, and the fit is performed independently on each data group; the uncertainty is given by the standard deviation of the best-fit parameter values.
The proximity of the best-fit values of transfer parameters $\alpha$ and $\kappa$ to the estimates in eqns.~(\ref{eqn:alpha}) and~(\ref{eqn:kappa}) validates the analysis of the apparatus design in Sec.~\ref{sec:D}.

\subsection{NMR response as a function of the Rabi frequency $\Omega_e$}
\noindent
In order to confirm the validity of our NMR model in the limit of small spin-tip angles, we record and analyze FID data for a range of excitation Rabi frequencies $\Omega_e$. For these measurements we keep the excitation pulse width at $20\uu{ms}$ -- approximately the coherence time of an axion-like dark matter field with Compton frequency near $40\uu{MHz}$.

We vary the Rabi frequency from $0.02\uu{rad/ms}$ to $0.88\uu{rad/ms}$. At each Rabi frequency, we apply 100 consecutive excitation pulses, spaced by $180\uu{ms}$. After each pulse, we sample the FID voltage, starting $5\uu{\mu s}$ after the end of the pulse, and lasting for $16.4\uu{\mu s}$.
We average the 100 FID data sets, and calculate the discrete Fourier transform $F[n]$ of the averaged FID, where index $n$ labels frequency points. Since we only sample the beginning of the FID, before it can start to decay, we model it as a sinusoidal signal at the excitation carrier frequency. We extract the amplitude of the spin ensemble transverse magnetization by numerically integrating the power spectrum $|F[n]|^2$ over a $400\uu{kHz}$-wide frequency band centered at the excitation carrier frequency, and using the pickup probe transfer function $\alpha$ to convert the voltage to magnetization. Uncertainties are calculated using bootstrapping: we group the 100 FID data sets into 5 sets of 20 and perform analysis on these 5 sets independently. Error bars are set at the standard deviation of the results for these 5 sets. To obtain the theory curve in Fig.~2(c) of main text, we use our Bloch equation model to generate numerical time-domain FID data, which we analyze in the same way as we analyze experimental data.

\section{Spectral properties of the CW NMR response}
\noindent
Under CW excitation with Rabi frequency $\Omega_e$ and carrier angular frequency $\omega_1$, the steady-state transverse magnetization of an unsaturated homogeneously-broadened spin ensemble is given by~\cite{Abragam1961}
\begin{align}
	M_1 = L(\omega_0-\omega_1)M_0\Omega_e T_2\cos{(\omega_1 t)},
	\label{eq:910}
\end{align}
where $M_0$ is the longitudinal magnetization, $T_2$ is the transverse coherence time, $\omega_0$ is the Larmor angular frequency, and $L$ is the Lorentzian lineshape function:
\begin{align}
	L(\omega_0-\omega_1)=\frac{1}{1+(\omega_0-\omega_1)^2T_2^2}.
	\label{eq:920}
\end{align}

Let us describe the spin ensemble inhomogeneous broadening with the excitation lineshape $h(\omega_0+\Delta)$, normalized such that
\begin{align}
	\int_{-\infty}^{\infty}h(\omega_0+\Delta)\,d\Delta=1.
	\label{eq:930}
\end{align}
Under CW excitation, the steady-state transverse magnetization is then
\begin{align}
	M_1 = uM_0\Omega_eT_2\cos{(\omega_1 t)},
	\label{eq:940}
\end{align}
where the spectral $u$ factor is given by the integral over the lineshape:
\begin{align}
	u = \int_{-\infty}^{\infty}L(\omega_0+\Delta-\omega_1)h(\omega_0+\Delta)\,d\Delta.
	\label{eq:950}
\end{align}

Let us estimate the value of $u$.
Our NMR measurements indicate that the excitation spectrum is much broader than $1/T_2$, therefore we can approximate the Lorentzian with the delta-function: $L(\omega_0+\Delta-\omega_1)\approx(\pi/T_2)\delta(\omega_0+\Delta-\omega_1)$. Furthermore, we approximate the excitation spectrum as a rectangular function, centered at $\omega_0$, with full width $\Gamma$ and height $1/\Gamma$. Then, provided $|\omega_0-\omega_1|<\Gamma/2$, we can approximate
\begin{align}
	u\approx \pi h(\omega_1)/T_2\approx \pi/(\Gamma T_2).
	\label{eq:960}
\end{align}

In order to more accurately determine $u$, we solved the Bloch equations with the experimentally-determined values $T_2 = (16.7 \pm 0.9)\uu{ms}$ and excitation spectrum with $\Gamma/(2\pi) = (78 \pm 2)\uu{kHz}$ (Fig. 2(b) in the main text). We obtained 
\begin{align}
	u = (3.8 \pm 0.3)\times 10^{-4},
	\label{eq:970}
\end{align}
in agreement with the estimate in Eq.~(\ref{eq:960}). 

The correction due to spin saturation by axion-like dark matter depends on the experimental sensitivity to the drive strength $\Omega_e$. Our signal detection threshold corresponds to $\Omega_e = 0.23\uu{rad/s}$, which corresponds to a 30\% correction to the value of the magnetization in Eq.~\eqref{eq:940}~\cite{Castner1959}. This correction was used for all our axion-like dark matter limits.

\section{Ferroelectric polarization of PMN-PT}
\begin{figure}[h]
	\includegraphics[width=0.9\textwidth]{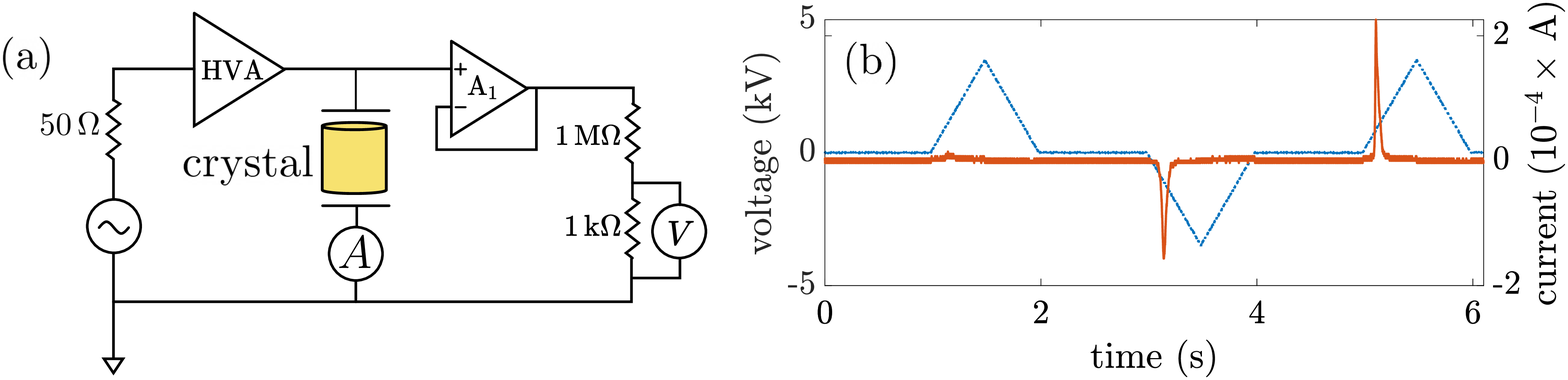}
	\caption{(a) Ferroelectric polarization setup, showing the signal generator controlling the high-voltage amplifier (HVA) that is connected to the electrodes in contact with the sample. TREK Model 610E high-voltage supply/amplifier/controller houses the HVA, as well as an ammeter $A$, a unity gain buffer amplifier $A_1$, and a voltmeter $V$. (b) Voltage applied at the output of HVA is measured at the voltmeter $V$ and converted to the voltage across the sample (blue dashed line). The current measured at the ammeter $A$ is plotted as the orange full line.}
	\label{fig:ferrorun}
\end{figure}

\noindent
We polarize the ferroelectric PMN-PT crystal by applying a voltage across its faces at room temperature. To ensure good electrical contact, we paint the faces with graphite paint, which is removed after polarization. We connect the crystal to the Trek model 610E-G-CE high-voltage amplifier as shown in Fig.~\ref{fig:ferrorun}(a). The amplifier measures the applied voltage and the current through the sample. In order to measure the ferroelectric hysteresis loop, we apply triangular voltage ramps with alternating polarities, Fig.~\ref{fig:ferrorun}(b). Current spikes are visible when the applied voltage is sufficient to reverse the ferroelectric polarization.
In this experimental run the crystal started with a remanent polarization corresponding to positive polarity, so there is no current spike during the first ramp. 
We obtain the sample polarization by integrating the current:
\begin{align}
	P(t) = \frac{q(t)}{\pi r_s^2} = \frac{1}{\pi r_s^2}\int_{0}^{t}{I(t') \d t'} ,
\end{align}
where $q(t)$ is the electric charge on the crystal surface and $r_s=2.3\uu{mm}$ is the base radius of the cylindrical sample. The hysteresis loop shown in Fig. 2(d) of the main text is the plot of polarization as a function of applied voltage. The remanent polarization $P_r$ persists after the voltage has been ramped down to zero. We verified that the remanent polarization does not decay after thermal cycling of the sample.

\section{Nuclear spin dynamics due to the EDM interaction with axion-like dark matter}

\subsection{P,T-odd axion-like dark matter physics}
\noindent
Axion-like cold dark matter is a classical field: $a(t)=a_0\cos{(\omega_a t)}$, where 
$\omega_a \approx m_ac^2/\hbar$. 
If the axion-field energy density dominates dark matter, then 
$\rho_{\text{DM}}=m_a^2a_0^2/2 \approx 3.6\times 10^{-42}\uu{GeV^4}$
~\cite{PDG2019}. In the QCD Lagrangian, this gives rise to an oscillating $\theta$ angle:
\begin{align}
	\theta(t) = \frac{a}{f_a}=\frac{a_0}{f_a}\cos{(\omega_a t)}.
	\label{eq:100}
\end{align}

Let us consider the nucleon EDM induced by axion-like dark matter:
\begin{align}
	d_n = g_d a = 2.4\times 10^{-16}\,\theta \uu{e\cdot cm} = 2.4\times 10^{-3}\,\theta \uu{e\cdot fm},
	\label{eq:110}
\end{align}
calculated with 40\% accuracy~\cite{Pospelov1999b,Graham2013}. Here $g_d$ is the EDM coupling constant~\cite{Graham2013}, introduced in the Lagrangian term:
\begin{align}
	\mathcal{L}_{\text{EDM}} = -\frac{i}{2}g_da\bar{\Psi}_N\sigma_{\mu\nu}\gamma_5\Psi_NF^{\mu\nu},
	\label{eq:120}
\end{align}
where $\Psi_N$ is the nucleon wavefunction, $F^{\mu\nu}$ is the electromagnetic field tensor, and $\sigma$ and $\gamma$ are the standard Dirac matrices.

From eqs.~(\ref{eq:100},\ref{eq:110}) we get the relationship between $g_d$ and $f_a$:
\begin{align}
	g_d = \frac{2.4\times 10^{-16}\uu{e\cdot cm}}{f_a} = \frac{3.6\times 10^{-3}\uu{GeV^{-1}}}{f_a},
	\label{eq:130}
\end{align}
where we used the natural unit conversions: $1\uu{cm}=5\times10^{13}\uu{GeV^{-1}}$ and $e=0.303$.

For the QCD axion, the decay constant is related to its mass:
\begin{align}
	m_a = 6\times10^{-10}\uu{eV}\,\left(\frac{10^{16}\uu{GeV}}{f_a}\right),
	\label{eq:140}
\end{align}
but for a generic ALP there is no such connection.

\subsection{Nuclear Schiff moments induced by the EDM coupling of axion-like dark matter}
\noindent
The nuclear Schiff moment~\cite{Schiff1963,Sandars1967,Sushkov1984,Flambaum2002} is defined as:
\begin{align}
	\bm{S} = \frac{e}{10}\left( \langle r^2\bm{r}\rangle - \frac{5}{3Z}\langle r^2\rangle\langle\bm{r}\rangle \right),
	\label{eq:200}
\end{align}
where $e$ is the elementary electric charge, $Z$ is the atomic number, and $\langle r^k \rangle = \int r^k\rho(\bm{r})d^3r$ are the integrals over nuclear charge density $\rho(\bm{r})$.
The Schiff moment sources the P- and T-odd electrostatic potential
\begin{align}
	\varphi(\bm{r}) = 4\pi(\bm{S}\cdot\nabla)\delta(\bm{r}).
	\label{eq:205}
\end{align}
Importantly, the definition of the Schiff moment in Ref.~\cite{Khriplovich1997} differs from this one by a factor of $4\pi$. We adopt the definition in Eq.~(\ref{eq:200}), noting the factor of $4\pi$ wherever we refer to Ref.~\cite{Khriplovich1997}.

The Schiff moment can be induced by a permanent EDM of a nucleon, or by P,T-odd nuclear forces~\cite{Khriplovich1997}. The contribution of P,T-odd nuclear forces is larger than the contribution of nucleon EDM~\cite{Sushkov1984}. Let us consider the two contributions separately, in the case of the $^{207}$Pb nucleus, whose ground state is $I^{\pi}=1/2^-$, having a neutron $3p_{1/2}$ hole in a closed-shell magic nucleus. 

\subsubsection{Nuclear Schiff moments induced by nucleon EDM}
\noindent
This contribution is due to non-coincident densities of nuclear charge and dipole moment. It can be estimated for $^{207}$Pb using Eq.~(8.76) in Ref.~\cite{Khriplovich1997}:
\begin{align}
	4\pi S_{\mathrm{EDM}} \approx d_n\times \frac{4\pi}{25}\frac{(K+1)I}{I(I+1)}r_0^2,
	\label{eq:210}
\end{align}
where $K=(\ell-1)(2I+1)=1$ and $r_0=1.25A^{1/3}=7.4\uu{fm}$. This estimate gives $S_{\mathrm{EDM}} \approx d_n\times 3\uu{fm^2}$.
More detailed calculations~\cite{Dzuba2002,Dmitriev2003a} give the result:
\begin{align}
	S_{\mathrm{EDM}} = d_n\times 1.9\uu{fm^2}.
	\label{eq:220}
\end{align}
In order to connect this to QCD axion physics, we use Eq.~(\ref{eq:110}):
\begin{align}
	S_{\mathrm{EDM}} = g_d a\times 1.9\uu{fm^2} = 5\times 10^{-3}\,\theta \uu{e\cdot fm^3}.
	\label{eq:240}
\end{align}

\subsubsection{Nuclear Schiff moments induced by P,T-odd nuclear forces}
\noindent
The P,T-odd nuclear interaction of a non-relativistic nucleon with nuclear core is parametrized by strength $\eta$~\cite{Sushkov1984}:
\begin{align}
	W = \frac{G_F}{\sqrt{2}}\frac{\eta}{2m}\bm{\sigma}\cdot\nabla \rho(\bm{r}),
	\label{eq:250}
\end{align}
where $G_F\approx 10^{-5}\uu{GeV^{-2}}$ is the Fermi constant, $m$ is the nucleon mass, $\bm{\sigma}$ is its spin, and $ \rho(\bm{r})$ is the density of core nucleons. A vacuum $\theta$ angle gives rise to this interaction via the P,T-odd pion-nucleon coupling constant~\cite{Khriplovich1997,Flambaum2014}: 
\begin{align}
	\eta = 1.8\times 10^6\,\theta.
	\label{eq:260}
\end{align}

Next we need to calculate the nuclear Schiff moment induced by the interaction~(\ref{eq:250}). Reference~\cite{Sushkov1984} states that the Schiff moment is suppressed by a factor $\sim 10$ for nuclei with a valence neutron, compared to a valence proton, and only core polarization leads to a non-zero effect. For example, the Schiff moment of $^{201}$Hg is estimated as $0.2\times 10^{-8}\eta\uu{e\cdot fm^3}$. However in Ref.~\cite{Flambaum1986} it was realized that virtual excitations in the core eliminate this suppression, and, in fact, the results for a valence neutron and a valence proton should be comparable. Here the Schiff moment of $^{201}$Hg is estimated as $2.4\times 10^{-8}\eta\uu{e\cdot fm^3}$, and the Schiff moment of $^{199}$Hg is estimated as $-1.4\times 10^{-8}\eta\uu{e\cdot fm^3}$.

The issue is complicated by nuclear many-body effects. These were numerically calculated for $^{199}$Hg in Refs.~\cite{Dmitriev2003,Ban2010}, giving a factor $\sim10$ reduction in the Schiff moment. However the physical origin of such a strong reduction is not clear. The only effect, not included in the shell model, that can change the value of the Schiff moment is the collective nuclear octupole deformation, and, if anything, that should increase the Schiff moment. Reference~\cite{Yanase2020} gives a result for $^{199}$Hg that is $\sim10\%$ away from the shell-model estimate. These authors attribute the Schiff moment suppression in Ref.~\cite{Ban2010} to the mixing with the $J^{\pi}=1/2^-_2$ state, for which they get a small Schiff moment value. However this small value itself is questionable. This state is an admixture of a soft quadrupole phonon ($J=2$) to the ground state, resulting still in $J=1/2$. The excited states do not have this quadrupole deformation, therefore the overlap matrix elements are likely to be small unless a lot of excited states are carefully taken into account. This  suggests that the calculation may have large intrinsic uncertainties. 

Importantly, $^{207}$Pb is close to a magic nucleus, which means that many-body effects should not play an important role here.  Therefore, until the many-body effects can be better understood, for $^{207}$Pb we retain the single-particle estimate of Ref.~\cite{Flambaum1986}:
\begin{align}
	S_{\eta} = 2\times 10^{-8}\eta\uu{e\cdot fm^3}=0.04\,\theta\uu{e\cdot fm^3}.
	\label{eq:270}
\end{align}
Note that this is a factor of eight larger than the result (13) in Ref.~\cite{Flambaum2020a}, where the $^{207}$Pb Schiff moment was taken to be the same as for the many-body suppressed $^{199}$Hg.
We can also see that this contribution is a factor of eight larger than the EDM contribution in Eq.~(\ref{eq:240}). We therefore neglect the EDM contribution, and use the above estimate (\ref{eq:270}).

Similar estimates were performed for $^{199}$Hg in Ref.~\cite{Stadnik2014}.

\subsection{Nuclear Schiff moment-induced spin energy shift in ferroelectric PMN-PT}
\noindent
The energy shift of each nuclear spin sublevel of a \chem{^{207}Pb^{2+}} ion in ferroelectric \chem{PbTiO_3} is estimated in Refs.~\cite{Mukhamedjanov2005,Ludlow2013}. The result of the full quantum chemistry calculation~\cite{Skripnikov2016} is:
\begin{align}
	\Delta\epsilon = 1.04\times 10^6\frac{x}{0.58\,\mbox{\AA}}\frac{S}{ea_B^3}\uu{[eV]}=1.2\times 10^{-8}\frac{x}{\mbox{\AA}}\frac{S}{\mathrm{e\cdot fm^3}}\uu{[eV]},
	\label{eq:300}
\end{align}
where $x$ is the displacement of the \chem{Pb^{2+}} ion with respect to the center of the oxygen cage, $S$ is the magnitude of the Schiff moment of the $^{207}$Pb nucleus, and $a_B=0.53\uu{\AA}$ is the Bohr radius. The nuclear spin is $I=1/2$; each of the two nuclear spin states shifts by this amount, in opposite directions. Since $\theta$ and $S$ exhibit sinusoidal time dependence, the experimentally relevant quantity is the Rabi angular frequency:
\begin{align}
	\Omega_a = \frac{1}{2}\frac{2\Delta\epsilon}{\hbar} = 1.8\times 10^7\frac{x}{\mbox{\AA}}\frac{S}{\mathrm{e\cdot fm^3}}\uu{[rad/s]}, 
	\label{eq:310}
\end{align}
where we used $\hbar = 6.58\times 10^{-16}\uu{eV\cdot s}$. We note that the spin driving field is ``linearly polarized'', and therefore the Rabi frequency contains an extra factor of $1/2$, which arises because only one of the two counter-rotating components of the linearly polarized drive is resonant (rotating wave approximation). 

Density functional theory calculations for PMN-PT give the Pb$^{2+}$ cation displacement from the center of the oxygen cage: $x_0=0.39$~\AA, and the average polarization: $P_0=55\uu{\mu C/cm^2}$~\cite{Grinberg2004}. Our experiment was performed with the crystal polarization  $P_r=22\uu{\mu C/cm^2}$, therefore we scale the average displacement to $x=0.16$~\AA. 

For $^{207}$Pb in ferroelectric PMN-PT we can use Eq.~(\ref{eq:270},\ref{eq:300}) and $x = 0.16$~\AA~to get:
\begin{align}
	\Delta\epsilon &= 8\times 10^{-11}\,\theta\uu{[eV]}, \\
	\Omega_a &= 1.2\times 10^5\,\theta\uu{[rad/s]}. 
	\label{eq:320}
\end{align}
To connect with the EDM $d_n$ and the coupling constant $g_d$, we use Eqs.~(\ref{eq:100},\ref{eq:110},\ref{eq:130}). For the energy shift we obtain
\begin{align}
	\Delta\epsilon =  d_n\uu{[e\cdot cm]}\times 3.4\times10^5\uu{[V/cm]}.
	\label{eq:325}
\end{align}
We can extract the effective electric field (which includes the Schiff screening factor~\cite{Budker2014}):
\begin{align}
	E^*=\Delta\epsilon/d_n=340\uu{[kV/cm]}.
	\label{eq:325}
\end{align}
For the drive Rabi frequency we obtain:
\begin{align}
	\Omega_a &= 1.2\times 10^5\frac{g_da_0}{3.6\times 10^{-3}\uu{[GeV^{-1}]}}\uu{[rad/s]} ,\\
	\hbar\Omega_a &=  2.2\times 10^{-17}(g_da_0)\uu{[GeV]},
	\label{eq:330}
\end{align}
where $g_d$ is in GeV$^{-2}$ and $a_0=\sqrt{2\rho_{\text{DM}}}/m_a$ is in GeV.
Let us introduce the sensitivity factor $\xi$, defined as $\hbar\Omega_a=\xi g_da_0$. Its estimated value is therefore
\begin{align}
	\xi =  2.2\times 10^{-17}\uu{[GeV^2]}.
	\label{eq:330}
\end{align}

There are several contributions to the theoretical uncertainty in $E^*$ and $\xi$. The uncertainty of the QCD calculations is $\approx 40\%$~\cite{Pospelov1999b,Graham2013}. The uncertainty of the solid-state calculation of the nuclear spin energy shift due to the Schiff moment is $\approx 30\%$~\cite{Mukhamedjanov2005,Ludlow2013,Skripnikov2016}. Therefore we estimate the total theoretical uncertainty in $E^*$ and $\xi$ at $\approx 50\%$.

\section{Nuclear spin dynamics due to the gradient interaction with axion-like dark matter}
\noindent
The non-relativistic Hamiltonian for the gradient interaction of spin $\vec{I}$ with axion-like dark matter field $a(\vec{r},t)$ is
\begin{align}
	H_{\text{aNN}} = g_{\text{aNN}}\vec{\nabla}a(\vec{r},t)\cdot\vec{I},
	\label{eq:330}
\end{align}
where $g_{\text{aNN}}$ is the coupling strength measured in units of GeV$^{-1}$, and we used natural units here $\hbar=c=1$~\cite{Graham2013,Garcon2019b}. In the first approximation we can write the axion-like dark matter field as: \begin{align}
	a(\vec{r},t)\approx a_0\cos{(\omega_at-\vec{k}\cdot\vec{r})},
	\label{eq:330}
\end{align}
where the field amplitude $a_0$ is fixed by the assumption that it dominates the dark matter energy density: $\rho_{\text{DM}} = m_a^2 a_0^2 / 2 = 3.6 \times 10^{-42}~\text{GeV}^4$~\cite{PDG2019, Graham2013}. 
We approximate the instantaneous value of the gradient $\vec{\nabla}a\approx m_a\vec{v}a$, where $\vec{v}$ is the instantaneous value of the velocity of the ALP field in the laboratory frame.
The Hamiltonian in natural units becomes:
\begin{align}
	H_{\text{aNN}} =(g_{\text{aNN}}a_0)m_a\vec{v}\cdot\vec{I}\cos{(\omega_at)}.
	\label{eq:330}
\end{align}
The product $g_{\text{aNN}}a_0$ is dimensionless, so we can restore the values of fundamental constants by dimensional analysis:
\begin{align}
	H_{\text{aNN}} =(g_{\text{aNN}}a_0)m_ac^2\frac{\vec{v}}{c}\cdot\vec{I}\cos{(\omega_at)}.
	\label{eq:330}
\end{align}
This interaction exerts a torque on nuclear spins, with the drive Rabi frequency given by 
\begin{align}
	\hbar\Omega_a =\frac{1}{2}(g_{\text{aNN}}a_0)m_ac^2\frac{v_{\perp}}{c},
	\label{eq:330}
\end{align}
where $v_{\perp}$ is the component of the velocity perpendicular to the direction of the leading field $B_0$.
As in the previous section, the spin driving field is ``linearly polarized'', and therefore the Rabi frequency contains an extra factor of $1/2$, which arises because only one of the two counter-rotating components of the linearly polarized drive is resonant (rotating wave approximation).

\section{Spectral properties of the spin response due to axion-like dark matter}
\noindent
In the first approximation we assume that the axion-like dark matter field is coherent, and drives the $^{207}$Pb nuclear spins at carrier angular frequency $\omega_a$ with Rabi frequency $\Omega_a$.
The steady-state transverse spin magnetization that develops under the action of this driving field is given by Eq.~(1) of the main text. The resulting voltage recorded by the ADC is:
\begin{align}
	V_a(t) = \alpha\mu_0M_a = \alpha u\mu_0M_0\Omega_aT_2\cos{(\omega_at)}.
	\label{eq:330}
\end{align}
The time-averaged power in this signal is 
\begin{align}
	\langle V_a^2\rangle = \frac{1}{2}(\alpha u\mu_0M_0\Omega_aT_2)^2.
	\label{eq:331}
\end{align}
Note that we use the term ``power'' in the signal processing context, and this is proportional to the physical power.

The Galactic axion-like dark matter halo field $a(t)$ is not perfectly coherent. 
In this work we search for the axion-like dark matter halo that follows the standard halo model~\cite{Turner1990,Evans2019}. In this model the ALP speeds $v$ in the Galactic frame follow the Maxwell-Boltzmann distribution
\begin{align}
	f_{\text{gal}}(v) = \frac{4v^2}{\sqrt{\pi} v_0^3} e^{-v^2 / v_0^2},
\end{align}
where $v_0 \approx 220~\text{km}/\text{s}$ is the most probable speed~\cite{Evans2019}. 
The laboratory frame moves relative to the Galactic frame with the average speed $v_{\text{lab}} \approx232~\text{km}/\text{s}$ which has annual and daily modulations due to, respectively, Earth's revolution about the Sun and Earth's rotation around its axis~\cite{Foster2018}.
The distribution of ALP speeds broadens the Fourier spectrum of the ALP field $a(t)$, giving it a characteristic linewidth $\approx v_0^2\nu_a/c^2\approx 10^{-6}\nu_a$. 
The power spectrum of the ALP field $a(t)$ is given by the function
\begin{align}
	f_0(\nu) = \frac{2c^2}{\sqrt{\pi} v_0 v_{\text{lab}} \nu_a} \exp{\left(-\frac{2c^2}{v_0^2} \frac{\nu - \nu_a}{\nu_a} - \frac{v_{\text{lab}}^2}{v_0^2}\right)} \sinh{\beta}, \label{eq:500}
\end{align}
where
\begin{align}
	\beta = \frac{2c v_{\text{lab}}}{v_0^2} \sqrt{\frac{2(\nu - \nu_a)}{\nu_a}}.
\end{align}
This spectral function is normalized so that
\begin{align}
	\int_{\nu_a}^{\infty}f_0(\nu) \,d\nu = 1.
	\label{eq:555}
\end{align}
This is the spectral lineshape used in searches for ALP-photon interactions~\cite{Du2018,Brubaker2017,Ouellet2019,Gramolin2020a}.

\subsection{EDM search}
\noindent
In our search for the ALP EDM interaction, the recorded voltage $V_{\text{EDM}}$ is directly proportional to the ALP field $a(t)$. Therefore the voltage power spectral density will have the same spectral shape $f_0(\nu)$. We make use of Parseval's theorem to ensure that the time-averaged power, Eq.~(\ref{eq:331}), matches the integral of the Fourier power spectrum, with the lineshape normalized as in Eq.~(\ref{eq:555}). The result is the expression for the voltage power spectrum:
\begin{align}
	V_{\text{EDM}}^2(\nu) = \frac{1}{2}(\alpha u\mu_0M_0\Omega_aT_2)^2f_0(\nu)
	= \frac{1}{2}(\alpha u\mu_0M_0T_2)^2\left(\frac{\xi g_d a_0}{\hbar}\right)^2f_0(\nu).
	\label{eq:330}
\end{align}

\subsection{Gradient search}
\noindent
In our search for the ALP gradient interaction, the recorded voltage $V_{\text{gr}}$ is proportional to the gradient of the ALP field, which includes the velocity of the ALP field in the lab frame. Therefore the voltage power spectrum has a different form:
\begin{align}
	f_1(\nu) = \frac{2c^2}{v_0^2 + v_{\text{lab}}^2 \sin^2{\zeta}} \frac{\nu - \nu_a}{\nu_a} \left[\sin^2{\zeta} + \frac{1}{\beta} \left(\coth{\beta} - \frac{1}{\beta}\right) \left(2 - 3\sin^2{\zeta}\right)\right] f_0(\nu),
	\label{eq:600}
\end{align}
where $\zeta$~is the angle between the vectors $\mathbf{B}_0$ and $\mathbf{v}_{\text{lab}}$. This spectral function is normalized so that
\begin{align}
	\int_{\nu_a}^{\infty}f_1(\nu) \,d\nu = 1.
\end{align}
A detailed analysis of the ALP velocity distribution in the laboratory reference frame, resulting in the ALP gradient spectral line shape~(\ref{eq:600}), will be published elsewhere.

Again, making use of Parseval's theorem to ensure that the time-averaged power equals the integral of the Fourier power spectrum, we write the expression for the voltage power spectrum:
\begin{align}
	V_{\text{gr}}^2(\nu) = \frac{1}{2}(\alpha u\mu_0M_0T_2)^2\left(\frac{g_{\text{aNN}} a_0 m_ac^2}{2\hbar}\right)^2\frac{v_0^2 + v_{\text{lab}}^2 \sin^2{\zeta}}{c^2}f_1(\nu).
\end{align}

\section{Data acquisition and analysis for the axion-like dark matter search}
\begin{figure}[h]
	\includegraphics[width=1\textwidth]{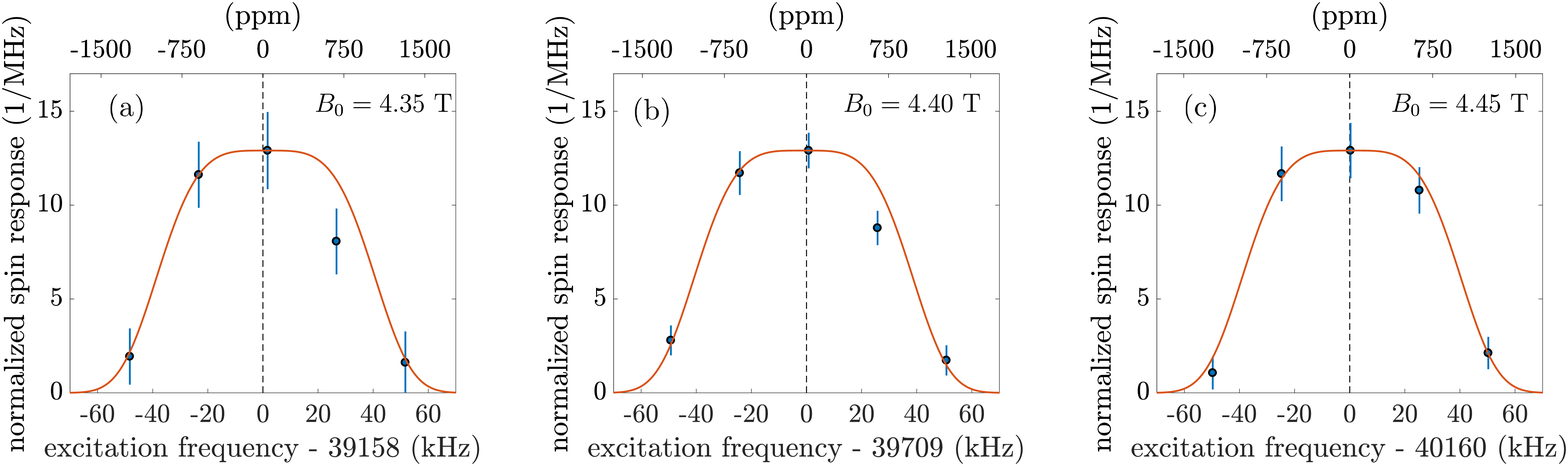}
	\caption{NMR calibration at the three values of the bias field $B_0$. FID data are recorded after excitation pulses at Rabi frequency $\Omega_e=0.88\uu{rad/ms}$ and pulse length $20\uu{ms}$. The excitation carrier frequency is plotted on the x-axis. Following the procedure used to obtain Fig. 2(b) in the main text, results are normalized so that the integral of the spectrum is unity.
		The error bars show one standard deviation uncertainties of the FID spectrum fits, performed as described in section~\ref{simbloch}.
		Each spectrum is modeled as a super-Gaussian of order 2 (Eq.~(\ref{eqn:gamma})) and constant width $78\uu{kHz}$ (orange line). The only free parameter is the central frequency. 
		The best-fit values of the central frequency for the three calibration data sets are: $\nu_0=(39159 \pm 1)\uu{kHz},\,\,(39708 \pm 1)\uu{kHz},\,\,(40160 \pm 2)\uu{kHz}$.
	}
	\label{fig:axion_nmr}
\end{figure}

\noindent
The experimental search for axion-like dark matter took place on October 7, 2019.
We varied the static magnetic field to sweep the spin Larmor frequency, starting at $40.16\uu{MHz}$ and ending at $39.16\uu{MHz}$, in $21$ steps with step size of $50\uu{kHz}$. This corresponds to magnetic fields between $B_0=4.45\uu{T}$ and $4.35\uu{T}$. During the recording of data sensitive to axion-like dark matter the superconducting magnet is in persistent mode with the power supply turned off and the excitation probe and cancellation coil are terminated with a $50\,\Omega$ resistor. 
Data are recorded at the ADC sampling rate of $250\uu{MS/s}$ and saved to a hard drive using the first-in, first-out mode of the ADC. At each value of $B_0$ we record $58\uu{s}$ of ``scan'' data, immediately followed by $58\uu{s}$ of ``re-scan'' data, analyzed as described below.
During the search we perform three pulsed NMR calibrations, at the first, the last, and the middle values of the magnetic field. Each calibration consists of FID data taken at five different excitation carrier frequencies near the corresponding Larmor frequency, Fig.~\ref{fig:axion_nmr}.

The data sensitive to axion-like dark matter are analyzed using Matlab on the Shared Computing Cluster, which is administered by Boston University’s Research Computing Services. The data-processing procedure consists of the following steps.
\begin{enumerate}[label=(\arabic*)]
	\item Divide data at each value of the magnetic field $B_0$ into 27 blocks. Each block contains $2^{29}$ points and corresponds to $2.15\uu{s}$ of data.
	The block duration exceeds the axion-like dark matter coherence time of $\approx 25\uu{ms}$ in the standard halo model.
	\item Perform the discrete Fourier transform on each data block without any windowing function, to obtain the spectral density $F[\nu]$. 
	Select the analysis frequency range between $39.1\uu{MHz}$ and $40.2\uu{MHz}$.
	The real and imaginary parts of the spectral density are Gaussian-distributed, therefore the power spectral density $|F[\nu]|^2=\operatorname{Re}(F[\nu])^2 + \operatorname{Im}(F[\nu])^2$ is a chi-square distribution with two degrees of freedom. Then, average $|F[\nu]|^2$ obtained from $27$ segments. Take a histogram and fit to a chi-square distribution and confirm that it has $54$ degrees of freedom. Repeat this for the scan (and separately, re-scan) data at each resonant frequency.
	\item \label{item:SG} Search for narrow RF interference spectral lines using the Savitzky-Golay filter with order 2 and length 31~\cite{Brubaker2017a}. 
	Spectral lines narrower than the ALP linewidth are distinguished by the difference between the filtered and raw power spectral densities. The points where this difference is above a threshold are marked as narrow spectral lines and are assigned the average value of their neighboring points.
	\item \label{item:lineshape} Optimally-filter data by convolving the power spectral density with the spectral lineshape for the ALP EDM interaction $f_0(\nu)$ given in Eq.~(\ref{eq:500}). The separation between distinct ALP search frequencies is set to the ALP signal linewidth $3(v_0^2/c^2)\nu_0/4$, where $\nu_0$ is the central Larmor frequency, determined by the value of the bias field $B_0$~\cite{Brubaker2017a,Foster2018}.
	\item Model the histogram of the optimally-filtered power spectral density with $100$ bins as the Gaussian distribution with mean $\mu$ and standard deviation $\sigma$. Calculate the detection threshold at $\mu+3.355\sigma$, corresponding to a $5\sigma$ detection with $95\%$ confidence level. Points above the threshold are ALP detection candidates. A detailed explanation of the choice of threshold value can be found in Refs.~\cite{Brubaker2017a,Gramolin2020a}.
\end{enumerate}

This analysis process is repeated for data taken at each of the $21$ settings of bias magnetic field $B_0$ in the scan. The spin response to an axion-like dark matter signal will only appear in the data set where $B_0$ is such that the ALP Compton frequency is within the magnetic resonance excitation spectrum.
For each data set we use the $80\uu{kHz}$ frequency band centered at Larmor frequency $\nu_0$, corresponding to the excitation spectrum, to search for the ALP signal, as described above. The rest of the spectral data within the $1\uu{MHz}$ scan range are used to reject residual background RF interference, which is not eliminated by the Savitzky-Golay filter. 
In addition, re-scan measurements are analyzed to eliminate statistical fluctuations, which are expected, given the large bandwidth of our search (look-elsewhere effect). The analysis procedure is as follows.
\begin{enumerate}[label=(\alph*)]
	\item At each value of bias magnetic field we consider $\approx 5000$ frequency points (independent values of the ALP Compton frequency). For Gaussian-distributed data we expect two points to be above the $3.355\sigma$ threshold. Typically we obtain $\approx 30$ candidates above the threshold. The excess candidates are due to RF interference.
	\item We compare candidate frequencies from the ``resonant'' data set (for which the frequency is within the excitation spectrum) to the candidate frequencies from the ``background'' data sets (for which the frequency is outside the excitation spectrum). If the candidate frequency appears in one of the background data sets, it is rejected as RF interference. On average this eliminates $\approx 28$ candidates at each value of $B_0$.
	\item We compare candidate frequencies from the scan and re-scan data sets. If a candidate frequency appears only in one of those data sets, it is rejected as a statistical fluctuation. On average this eliminates $\approx 2$ candidates at each value of $B_0$.
\end{enumerate}
This analysis procedure rejects all candidates above the $3.355\sigma$ threshold at all values of $B_0$. We do not detect an axion-like dark matter signal. 
Therefore, for each value of $B_0$, we quote the $g_d$ coupling value that corresponds to the $5\sigma$ value of the power spectral density as the 95\% confidence interval limit~\cite{Gramolin2020a}.

\begin{figure}[h]
	\includegraphics[width=0.35\textwidth]{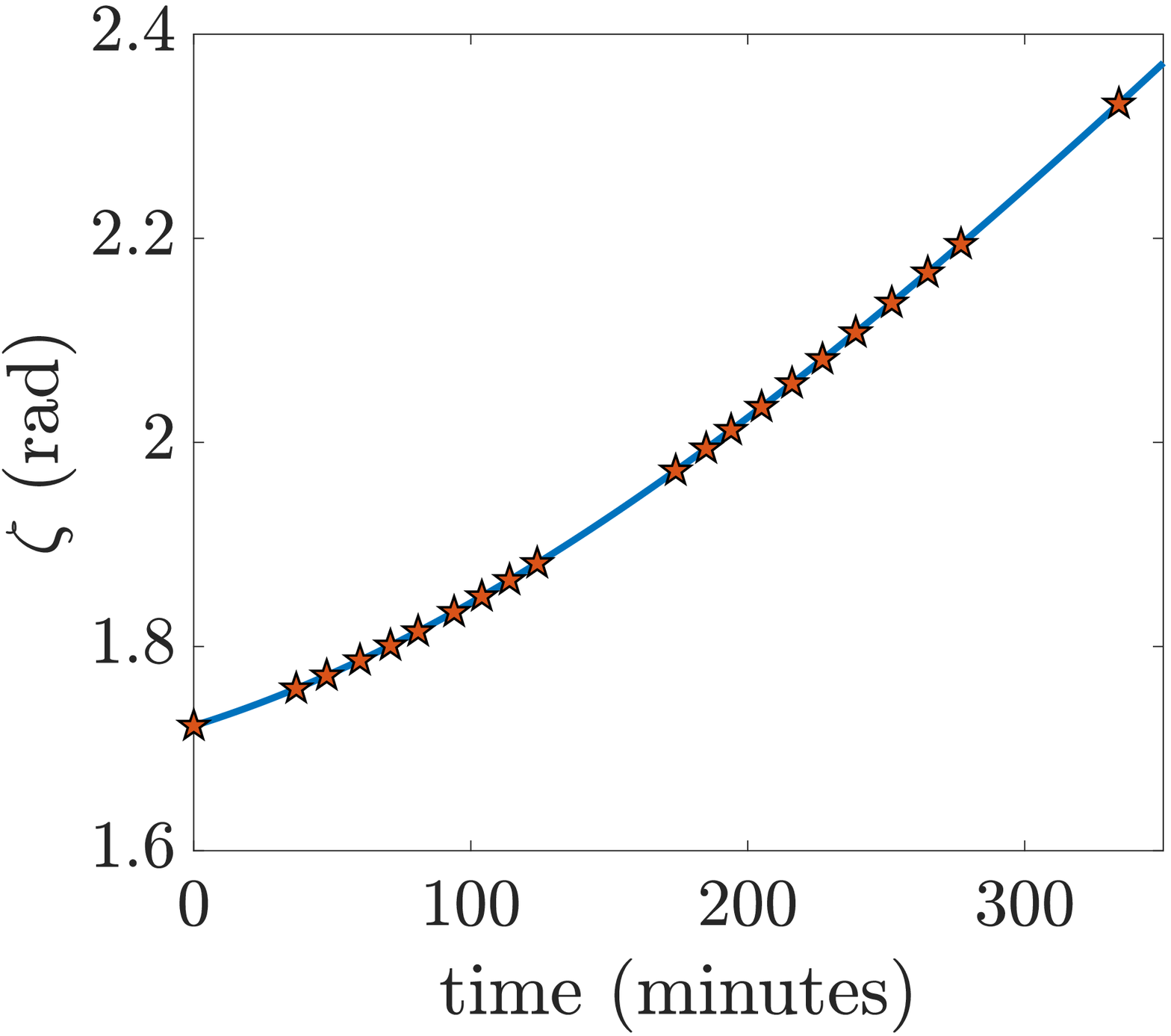}
	\caption{Angle $\zeta$ between the bias magnetic field $\mathbf{B}_0$ and the laboratory velocity vector $\mathbf{v}_{\text{lab}}$ as a function of time offset from the start of the experimental search for axion-like dark matter at 18:41 UTC on October 7, 2019 to 00:30 UTC on October 8, 2019. Stars indicate the times at which data are recorded for different values of $B_0$. The magnitude of the laboratory velocity is $v_{\text{lab}} = 226.5 \pm 0.5~\text{km}/\text{s}$ for the entire duration of data taking. The velocity $\mathbf{v}_{\text{lab}}$ is calculated for the Physics Department at Boston University ($42\degree20'53.8"\mathrm{N}, 71\degree06\mathrm{'}01.8"\mathrm{W}$) using the Python code~\cite{tassle2020} based on the Astropy library~\cite{astropy:2013,astropy:2018}.}
	\label{fig:zeta}
\end{figure}

We search for the gradient coupling $g_{\text{aNN}}$ of axion-like dark matter using the same steps as described above, with the standard halo model lineshape in step~\ref{item:lineshape} replaced by the gradient coupling lineshape $f_1(\nu)$, given in Eq.~(\ref{eq:600}). We calculate the angle $\zeta$ at each value of $B_0$ during the scan, based on the coordinates of our laboratory and the time at which the data are recorded, Fig.~\ref{fig:zeta}. 

Our analysis for the gradient coupling $g_{\text{aNN}}$ rejects all candidates above the $3.355\sigma$ threshold at all values of $B_0$. Therefore, for each value of $B_0$, we quote the $g_{\text{aNN}}$ coupling value that corresponds to the $5\sigma$ value of the power spectral density as the 95\% confidence interval limit. We note that the variation in $\zeta$ throughout the scan means that the shape of the limit curves for $g_d$ and for $g_{\text{aNN}}$ is slightly different in Fig.~4(b) of the main text, however this difference is smaller than the line thickness on the logarithmic plot.

\subsection{Testing the data analysis procedure by injecting ALP signals}
\begin{figure}[h]
	\includegraphics[width=0.7\textwidth]{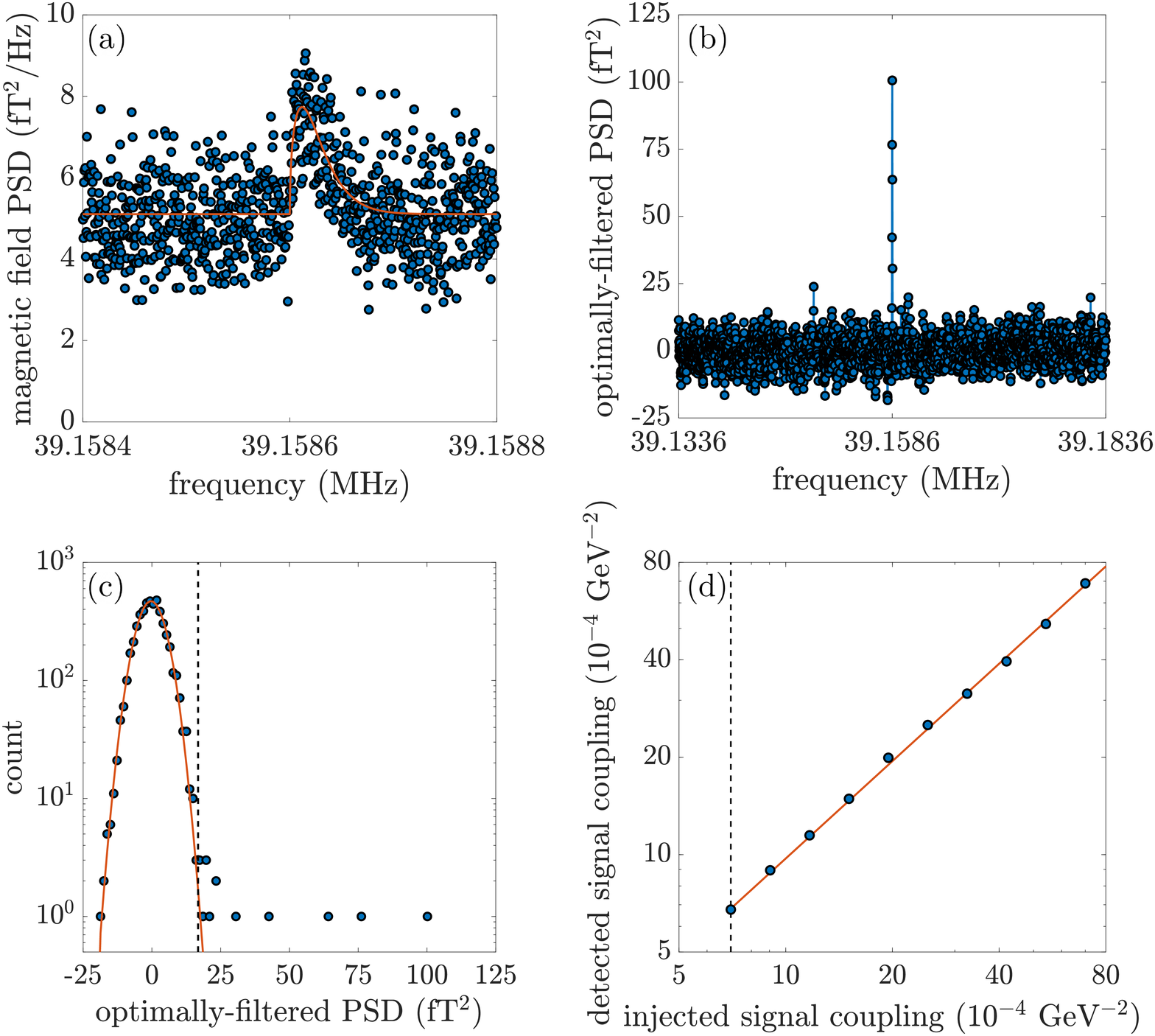}
	\caption{Injecting simulated axion-like dark matter signals into experimental data. 
		(a) A $400\uu{Hz}$-wide band of experimental power spectrum, with an injected signal at Compton frequency $\nu_a=39.1586\uu{MHz}$ and with magnetic field PSD of $2.6\uu{fT^2/Hz}$. Experimental data are shown as blue circles, the injected signal is shown as the orange line.
		(b) The optimally-filtered spectrum within the $50\uu{kHz}$ frequency band centered on $\nu_a$. 
		(c) The histogram of optimally-filtered PSD data (blue circles) within the $80\uu{kHz}$ band centered on the Larmor frequency $\nu_0=39.16\uu{MHz}$. Data are sorted into 100 bins, and the Gaussian fit is shown as the orange line. The $3.355\sigma$ detection threshold (vertical dashed black line) is at $17\uu{fT^2}$. 
		(d) Recovered coupling for injected signals with coupling strengths varying logarithmically from $g_d = 7.0\times10^{-4}\uu{GeV^{-2}}$ to $g_d = 7.0\times10^{-3}\uu{GeV^{-2}}$ at different Compton frequencies, sampled randomly between $\nu_a = 39.1185\uu{MHz}$ and $\nu_a = 39.1985\uu{MHz}$. The orange line shows the linear fit, from which we extract the $(2.7 \pm 0.8)\%$ signal suppression.}
	\label{fig:axion_inj}
\end{figure}

\noindent
We test our data-analysis procedure by injecting into the experimental spectra synthetic axion-like dark matter signals with the lineshape given by Eq.~(\ref{eq:500}). Figure \ref{fig:axion_inj}(a) shows the spectrum with an injected signal at Compton frequency $\nu_a=39.1586\uu{MHz}$ and with magnetic field PSD of $2.6\uu{fT^2/Hz}$. After optimal filtering, the injected signal shows up as a candidate with amplitude $101\uu{fT^2}$, as shown in Fig.~\ref{fig:axion_inj}(b). The histogram of the optimally-filtered data points shows that this injected signal is detected at $20\sigma$ significance, Fig.~\ref{fig:axion_inj}(c).
We test the recovery of the coupling strength by injecting 10 simulated signals, whose coupling strength is varied between $g_d = 7.0\times10^{-4}\uu{GeV^{-2}}$ and $g_d = 7.0\times10^{-3}\uu{GeV^{-2}}$ and whose Compton frequencies are selected randomly between $\nu_a = 39.1185\uu{MHz}$ and $\nu_a = 39.1985\uu{MHz}$. The coupling strengths recovered from detected signals are shown in Fig.~\ref{fig:axion_inj}(d). We find that, on average, our analysis procedure results in a $(2.7 \pm 0.8) \%$ suppression in the recovered coupling strength. This is due to the discrete sampling of the ALP search frequencies. If the injected ALP frequency falls between the search frequencies, there is a small mismatch in the lineshapes, which reduces the recovered coupling strength.
The limits reported in the main text are corrected for this suppression.

\subsection{Projected sensitivity reach}
\noindent
Our experimental results demonstrate the feasibility of using solid-state nuclear magnetic resonance to search for axion-like dark matter. There are several bounds on the relevant interactions of axion-like dark matter in this mass range, based on analysis of cooling dynamics of supernova SN1987A~\cite{Raffelt2008,Budker2014,Graham2015a}, and of Big-Bang nucleosynthesis~\cite{Blum2014}. However these model-dependent bounds are subject to significant caveats and uncertainties, and may be evaded altogether~\cite{DeRocco2020,Bar2020}.
Stringent experimental limits on $g_d$ and $g_{\text{aNN}}$ exist at much lower ALP masses~\cite{Vasilakis2009,Abel2017a,Wu2019a,Garcon2019b,Terrano2019,Roussy2020}, but the mass range probed in the current search has been, until now, experimentally unexplored.

\begin{figure}[h]
	\includegraphics[width=1\textwidth]{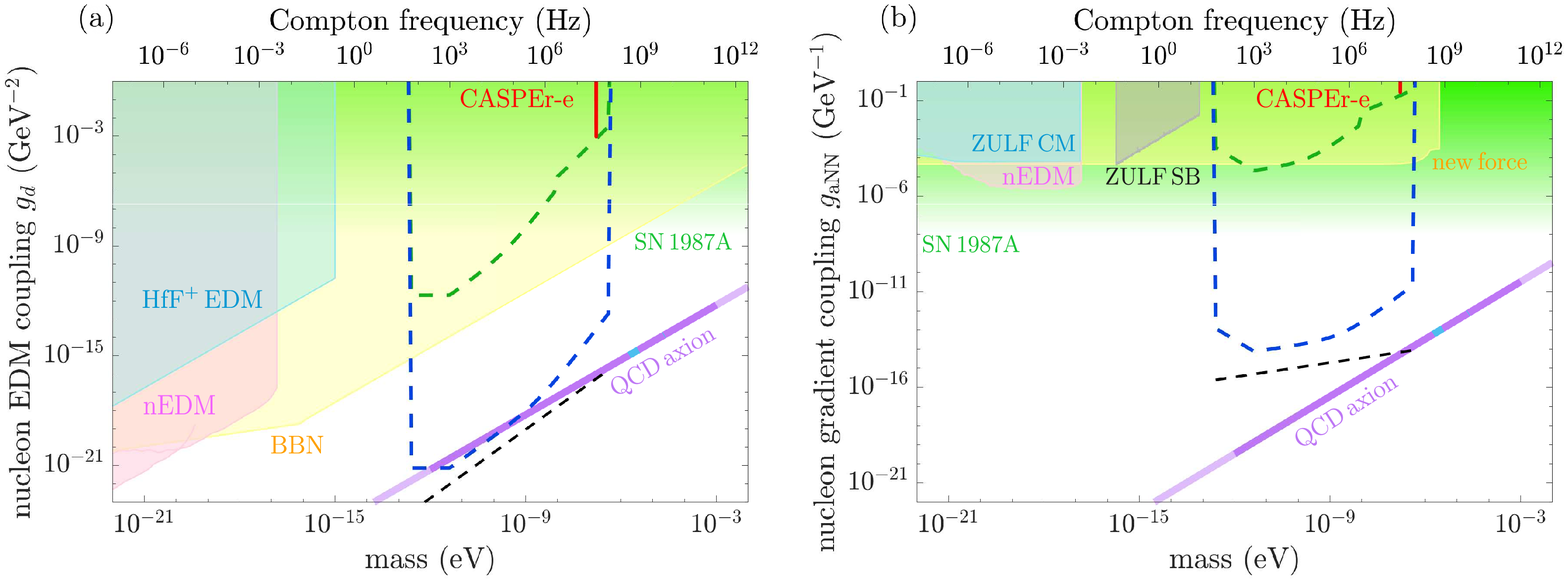}
	\caption{Existing bounds and sensitivity projections for the:
		(a) EDM and (b) gradient coupling of axion-like dark matter. 
		The region shaded in red is the exclusion at 95$\%$ confidence level placed by this work (CASPEr-e).
		The purple line shows the QCD axion coupling band. The darker purple color shows the mass range motivated by theory~\cite{Graham2013}. The blue regions mark the mass ranges where the ADMX and HAYSTAC experiments have probed the QCD axion-photon coupling~\cite{Du2018,Brubaker2017}.
		The green region is excluded by analysis of cooling in supernova SN1987A, with color gradient indicating theoretical uncertainty~\cite{Graham2013}.
		The dashed green line marks the projected $5\sigma$ sensitivity of our CASPEr-e search with a $4.6\uu{mm}$ sample, as used in current work. The dashed blue line marks the projected $5\sigma$ sensitivity of our CASPEr-e search with an $80\uu{cm}$ sample, operating at $100\uu{mK}$ temperature. Implementing a resonant coupling circuit will enable operation with a smaller sample. The black dashed line marks the sensitivity limited by the quantum spin projection noise~\cite{Budker2014}. This is sufficient to detect the EDM coupling of the QCD axion across the 6-decade mass range from $\approx0.3\uu{peV}$ to $\approx500\uu{neV}$.
		The other bounds are as follows.
		(a) The pink region is excluded by the neutron EDM (nEDM) experiment~\cite{Abel2017a}.
		The blue region is excluded by the HfF$^+$ EDM experiment~\cite{Roussy2020}.
		The yellow region is excluded by analysis of Big Bang nucleosynthesis (BBN)~\cite{Blum2014}.
		(b) The pink region is excluded by the neutron EDM (nEDM) experiment~\cite{Abel2017a}.
		The blue region is excluded by the zero-to-ultralow field comagnetometer (ZULF CM) experiment~\cite{Wu2019a}.
		The gray region is excluded by the zero-to-ultralow field sideband (ZULF SB) experiment~\cite{Garcon2019b}.
		The yellow region is excluded by the new-force search with K-$^3$He comagnetometer~\cite{Vasilakis2009}.
		The bounds are shown as published, although corrections should be made to some of the low-mass limits, due to stochastic fluctuations of the axion-like dark matter field~\cite{Centers2019}.
	}
	\label{fig:global_lims}
\end{figure}

The current sensitivity is not yet sufficient to reach the benchmark QCD axion level. The two main reasons are: (1) the CSA-induced inhomogeneous broadening of the NMR linewidth of the $^{207}$Pb nuclear spin ensemble, and (2) the small size of our PMN-PT sample. We plan to circumvent the inhomogeneous broadening by concentrating our future searches on the lower Compton frequencies ($\nu_a<1\uu{MHz}$), where the linewidth will be dominated by the $T_2$ spin coherence time, rather than CSA. The long $T_1$ relaxation time will allow us to pre-polarize the nuclear spins, retaining their polarization even at lower fields. We plan to use Superconducting Quantum Interference Devices (SQUIDs) to detect the transverse magnetization in this frequency range. The green dashed curves in Fig.~\ref{fig:global_lims} show the projected experimental sensitivity for the search with the same $4.6\uu{mm}$ sample as used in the current work. The cutoff at the low frequency end is set at the $1/T_2$ NMR linewidth, and the cutoff at high frequencies is set by the Larmor frequency at the maximum magnetic field of $15\uu{T}$.

In order to reach sufficient sensitivity to probe the QCD axion coupling strengths, we plan to scale up the volume of the ferroelectric sample. If the sample is coupled to the SQUID sensor with a broadband circuit, 
sample size of $\approx80\uu{cm}$ and operation at $\approx100\uu{mK}$ temperature are sufficient to reach the QCD axion line over $\approx 3$ decades in mass, Fig.~\ref{fig:global_lims}, blue dashed line. Implementing a resonant coupling circuit with a modest quality factor $\approx 1000$ may allow us to reach this sensitivity level with a sample that is an order of magnitude smaller. The ultimate sensitivity limit is determined by the nuclear spin projection noise, Fig.~\ref{fig:global_lims}, black dashed line.

